\documentclass[a4paper,fleqn,review]{cas-sc}

\usepackage{hyperref}

\usepackage{textgreek}
\usepackage{amsthm}
\usepackage{tabularx}

\usepackage[numbers]{natbib}
\usepackage{amsthm,amssymb,tikz-cd}

\usepackage{makecell}
\usepackage{placeins}

\usepackage[dvipsnames]{xcolor}

\usepackage{soul}

\usepackage{makecell}
\usepackage{pifont}
\usepackage{graphicx}

\usepackage{tikz}
\usetikzlibrary{positioning}
\usetikzlibrary{shapes.geometric}

\makeatletter
\renewcommand{\fnum@figure}{Fig. \thefigure}
\makeatother

\theoremstyle{definition}

\theoremstyle{plain}

\usepackage{xcolor}

\usepackage{fancyhdr} 
\makeatletter
\def\ps@first{%
   \let\@oddhead\@empty
   \let\@evenhead\@empty
   \def\@oddfoot{}
   \let\@evenfoot\@oddfoot
}
\usepackage{amssymb}

\begin{document}

\title[mode=title]{A Comprehensive Incremental and Ensemble Learning Approach for Forecasting Individual Electric Vehicle Charging Parameters}

\nonumnote{Preprint. Under review.}

\shortauthors{P. Alikhani et~al.}

\author[1]{\textcolor{black}{Parnian Alikhani*}\corref{cor1}}[type=editor,auid=000,bioid=1,orcid=0000-0003-1678-1588]
\ead{p.alikhani@uu.nl}
\credit{Conceptualization, Methodology, Validation, Visualization, Writing - original draft, Writing - review and editing}
\cortext[cor1]{Corresponding author}
\author[1]{\textcolor{black}{Nico Brinkel}}[type=editor,
                        auid=000,bioid=1,orcid=0000-0001-9973-2890]
\credit{Conceptualization, Methodology, Visualization, Validation, Writing - review and editing}
\author[2]{\textcolor{black}{Wouter Schram}}[type=editor,
                        auid=000,bioid=1,orcid=0000-0003-3407-7893]
\credit{Conceptualization, Methodology, Visualization, Validation, Writing - review and editing}
\author[1]{\textcolor{black}{Ioannis Lampropoulos}}
\credit{Validation, Writing - review and editing, Supervision, Funding acquisition}
\author[1]{\textcolor{black}{Wilfried van Sark}\corref{}}
\credit{Validation, Writing - review and editing, Supervision, Funding acquisition}

\affiliation[1]{organization={Copernicus Institute of Sustainable Development, Utrecht University},
addressline={Princetonlaan 8A},
postcode={3584 CB},
city={Utrecht},
country={The Netherlands}}
\affiliation[2]{organization={Department of Design, Production, and Management, University of Twente},
addressline={Drienerlolaan 5},
postcode={7500 AE},
city={Enschede},
country={The Netherlands}}

\begin{abstract}
Electric vehicles (EVs) have the potential to reduce grid stress through smart charging strategies while simultaneously meeting user demand. This requires accurate forecasts of key charging parameters, such as energy demand and connection time. Although previous studies have made progress in this area, they have overlooked the importance of dynamic training to capture recent patterns and have excluded EV sessions with limited information, missing potential opportunities to use these data. To address these limitations, this study proposes a dual-model approach incorporating incremental learning with six machine-learning models to predict EV charging session parameters. This approach includes dynamic training updates, optimal features, and hyperparameter set selection for each model to make it more robust and inclusive. Using a data set of 170,000 measurements from the real world electric vehicle session, week-long charging parameters were predicted over a one-year period. The findings reveal a significant difference between workplace and residential charging locations regarding connection duration predictability, with workplace sessions being more predictable. The proposed stacking ensemble learning method enhanced forecasting accuracy, improving $R^2$ by 2.83\% to 43.44\% across all parameters and location settings. A comparison of the two models reveals that incorporating user IDs as a feature, along with the associated historical data, is the most significant factor influencing the accuracy of the forecast. Forecasts can be used effectively in smart charging and grid management applications by incorporating uncertainty quantification techniques, allowing charge point operators to optimize charging schedules and energy management.

\end{abstract}

% Keywords
% Each keyword is separated by \sep

\maketitle

\section{Introduction}
\subsection{Problem definition}
The adoption of electric vehicles (EVs) is accelerating worldwide, with global sales reaching 14 million units in 2023, representing 18\% of total vehicle sales \cite{IEA}. This rapid transition to EVs offers an immediate reduction in $\mathrm{CO_2}$ emissions and contributes substantially to improved air quality, particularly in urban areas \cite{ercan2022autonomous}. However, this shift also introduces a pressing need for expanded EV charging infrastructure to support increasing demand. Without careful energy demand management, the growth in electric vehicle charging demand can challenge the stability of the electricity grid, causing voltage fluctuations, grid congestion, and heightened grid reinforcement costs \cite{sayed2022electric}. To maintain stability and address potential challenges arising from increased EV adoption, grid operators increasingly focus on deploying flexibility options such as smart charging and vehicle-to-grid (V2G) technologies. Smart charging has emerged as a key solution to alleviate grid stability problems by dynamically scheduling the charging demand of EV charging sessions to reduce peak load \cite{li2024comparative}. In addition, smart charging can be used for other functions, for instance, by aligning demand with low-price hours \cite{brinkel2023dynamic} or periods of increased renewable energy generation \cite{meng2024revolutionizing}. 

One of the main problems limiting the effective implementation of EV smart charging is the high variability and unpredictability of EV charging session characteristics \cite{gschwendtner2024leveraging}. This variability is particularly critical for Charge Point Operators (CPOs), who are responsible for managing charging infrastructure and ensuring reliable service. To effectively deploy smart charging, CPOs need accurate insights into EV energy demand and departure times. However, these aspects are not communicated in the most widely used versions of EV charging communication protocols. The misestimation of these aspects can create two key challenges. If smart charging is applied too optimistically, a user departing earlier than expected may experience unmet charging demand. In contrast, an overly cautious approach to preventing user dissatisfaction can result in the inefficient application of smart charging. 

To address this, CPOs can leverage forecasting to manage the uncertainties associated with individual charging sessions and use the insights to deploy effective real-time smart charging strategies. Forecasting these charging session characteristics accurately requires models that can capture the nuances of individual EV charging sessions. Accurate forecasting of EV session parameters is essential for addressing the practical challenges of smart charging implementation, such as coordinating charging demand with grid constraints and ensuring user convenience while optimizing operational efficiency, minimizing environmental impact \cite{BRINKEL2020115285}, or maximizing economic viability \cite{xiong2017realtimebidirectionalelectricvehicle}. By enabling effective load distribution and adaptive charging strategies, more accurate forecasts help mitigate grid congestion, reduce unnecessary infrastructure investments, and enhance overall system reliability \cite{fescioglu2023electric}.

\subsection{Literature review}
In the context of EV charging demand, time-series forecasting has emerged as a critical area of research, leading to the development of numerous forecasting methods. Yi et al. \cite{yi2022electric} propose a time-series forecasting model for predicting monthly commercial EV charging demand through a deep learning approach known as Sequence to Sequence (Seq2Seq). This model demonstrates significant improvements over other models in multi-step forecasting tasks, emphasizing its effectiveness in handling sequential data. Similarly, Kim and Kim \cite{en14051487} focus on forecasting daily energy consumption by leveraging historical charging data, weather conditions, and day-of-the-week effects. Their approach combines statistical models, such as the Autoregressive Moving Average (ARMA) and Autoregressive Integrated Moving Average (ARIMA), with deep learning techniques like Long Short-term Memory (LSTM), incorporating both past values and external factors. In addition, Van Kriekinge et al. \cite{wevj12040178} employ LSTM networks to predict day-ahead EV charging demand at 15-minutes intervals. Furthermore, Rathore et al. \cite{rathore2023prediction} explore a range of machine learning models, including Random Forest (RF), XGBoost, and neural network architectures, to predict EV energy demand based on historical charging data. Their findings reveal that the RF and XGBoost models offer the highest forecast accuracy.

While these studies advance forecasting at aggregated levels, optimizing smart charging requires a more granular approach focusing on individual EV charging sessions. Predicting session-specific parameters such as energy demand per session and connection duration, rather than broader time-series trends, enables greater responsiveness to real-time grid conditions. Frendo et al. \cite{frendo2020improving} forecast the time of day an EV is unplugged, but their model assumes that the vehicle is always unplugged on the same day it was initially plugged in. This assumption, however, does not hold for residential charging scenarios, where overnight charging is prevalent and often extends the unplugging time to the following day. Phipps et al. \cite{phipps2023customized} include forecasts for parking durations ranging from two to twenty-four hours, suggesting that only these sessions are relevant because shorter sessions offer limited flexibility. However, this approach overlooks the challenge of determining whether a session will be particularly long or short at the time the EV is initially connected to the charging station.

Lee et al. \cite{lee2019acn} focus on predicting both the duration and energy consumption for individual charging sessions. However, their evaluation is limited to analyzing the overall test error distribution, without examining variations in forecast accuracy across different vehicles. Table \ref{tab:lit} provides a summary of related works in the literature, primarily focusing on studies that address EV charging parameters for undifferentiated or single-location settings. Since EV charging parameters are highly dependent on location, comparing forecast results across two distinct settings offers valuable insights into the feasibility of each approach and highlights key features relevant to different environments.

EV charging patterns are highly variable, with influencing factors including location, time of day, and user behavior, making accurate forecasting challenging. Single-model approaches often exhibit substantial forecast errors due to this behavioral variability. Ensemble learning methods, which combine multiple models to leverage their unique strengths, have shown promise in enhancing forecast accuracy across various domains \cite{9319916, BUZNA2021116337}. This study applies ensemble learning to accommodate the behavioral diversity in EV charging patterns, aiming to reduce error variances across user scenarios.

Identifying the most relevant input features is crucial for accurate EV demand forecasting. Previous studies have shown that effective feature selection can significantly improve model performance by isolating the most influential variables \cite{BRINKEL2023100297, biom10030454}. Additionally, the availability of historical experimentally determined data on EV charging has proven essential in enhancing model performance, with recent data updates further improving forecast accuracy by reflecting current trends and habits \cite{hewamalage2023forecast, netzell2023applied}. By examining the relationship between forecast performance and data volume, this study provides insights into the optimal data requirements for reliable forecasts, supporting a balance between data storage, computational efficiency, and forecast precision.

\begin{table}[!b]
\centering
\caption{Overview of relevant literature on forecasting parameters for individual EV charging sessions.} 
\label{tab:lit}
\resizebox{\linewidth}{!}{ % Resize table to fit within line width
\begin{tabular}{c c c c c c c c c}
\hline
\multicolumn{1}{c}{Paper} & \multicolumn{2}{c}{EV parameters for forecasting} & \multicolumn{1}{c}{\makecell{Station \\location type}} & \multicolumn{1}{c}{\makecell{Forecasting method*}} & \multicolumn{1}{c}{Considered features} & \multicolumn{1}{c}{\makecell{Optimal \\feature \\selection}} & \multicolumn{1}{c}{\makecell{Dynamic training \\updates}} & \multicolumn{1}{c}{\makecell{Forecasting \\ analysis}} \\ 
\cline{2-3}
 & \makecell{Charging \\ energy demand} & \makecell{EV connection \\ duration (departure)} & & & & & & \\ 
\hline

\cite{frendo2020improving} & \ding{55} & \ding{51} & Workplace & \makecell{LR, XGBoost, \\ANN} & \makecell{Limited Temporal, \\historical} & \ding{55} & \ding{55} & \makecell{Performance of models, \\ feature importance} \\ \hline

\cite{phipps2023customized} & \ding{55} & \ding{51} & Spatial clustering & \makecell{BRR, GPR,\\NGBoost, QRNN} & \makecell{Temporal, \\ limited historical} & \ding{55} & \ding{55} & \makecell{Performance of models, \\ uncertainty quantification} \\ \hline

\cite{lee2019acn} & \ding{51} & \ding{51} & Workplace & \makecell{GMM} & \makecell{Temporal, \\historical} & \ding{55} & \ding{55} & \makecell{Performance of models} \\ \hline

\cite{straka2022role} & \ding{55} & \ding{51} & Undifferentiated & \makecell{QR, GBRT,\\ANN} & \makecell{Limited temporal, \\ historical} & \ding{55} & \ding{55} & \makecell{Performance of models} \\ \hline

\cite{cai2022optimizing} & \ding{51} & \ding{51} & Undifferentiated & \makecell{QR, GBQR,\\ Quantile RF} & \makecell{Limited temporal, \\ historical} & \ding{55} & \ding{55} & \makecell{Performance of models, \\ feature importance} \\ \hline

\cite{KREFT2024123544} & \ding{51} & \ding{51} & Residential & \makecell{LR, QR,\\HGBR} & \makecell{Temporal, \\historical} & \ding{55} & \ding{55} & \makecell{Performance of models, \\ feature importance} \\ \hline

\cite{shahriar2021prediction} & \ding{51} & \ding{51} &  Public & \makecell{RF, SVR, \\XGBoost, ANN,\\ Ensemble learning} & \makecell{Temporal, \\historical,\\meteorological} & \ding{51} & \ding{55} & \makecell{Performance of models, \\ feature importance} \\ \hline

This study & \ding{51} & \ding{51} & \makecell{Residential \\and workplace} & \makecell{LR, SVR, DT,\\ RF, XGBoost,\\ Ensemble learning} & \makecell{Temporal, \\historical,\\meteorological} & \ding{51} & \ding{51} & \makecell{Performance of models\\ across different EV\\ fleet locations, \\feature importance, \\training dataset \\size impact} \\ 
\hline
\end{tabular}}
\parbox{\textwidth}{\scriptsize *LR: Linear Regression, BRR: Bayesian Ridge Regression, GPR: Gaussian Process Regression, QR: Quantile Regressor, SVR: Support Vector Regression, DT: Decision Tree, RF: Random Forest, GMM: Gaussian Mixture Model, GBRT: Gradient Boosted Regression Trees, XGBoost: Extreme Gradient Boosting, NGBoost: Natural Gradient Boosting, HGBR: Histogram-based Gradient Boosting Regression, ANN: Artificial Neural Network, QRNN: Quantile Regression Neural Network}
\end{table}

\subsection{Contributions}
Despite advancements in forecasting EV charging demand, a gap remains in the accuracy of session-specific forecasts. Forecasting individual EV charging session characteristics — rather than aggregated demand — enhances the ability of  CPOs to improve their smart charging optimization approaches. As summarized in Table \ref{tab:lit}, previous studies in the literature have effectively utilized machine learning to predict session duration and energy consumption. However, they have largely been limited to single or undifferentiated location settings and have often neglected incremental learning approaches and focused instead on static or batch-based methods that do not adapt continuously to new data. Moreover, there is limited insight into feature importance, particularly user-specific features, in forecasting models. Most studies also lack feature selection, potentially incorporating irrelevant features that degrade model performance rather than optimizing it. This highlights the importance of both feature selection for improved predictive accuracy and incremental learning, which enables the model to adapt to new data from an updated training set, ensuring ongoing relevance and accuracy.

This study presents a comprehensive approach for forecasting charging session-specific parameters, in which we systematically compare various forecast models, including advanced machine learning and ensemble techniques, to identify optimal methods for forecasting per-session energy demand and connection duration. Additionally, our method integrates training data updates, optimal feature selection, and user-specific dual-model approaches to enhance forecast accuracy, ensuring the model remains adaptable and relevant as data evolves.

The contributions of this work can be summarized as follows:

\begin{enumerate}
    \item A comparative analysis of forecast models for session-specific EV metrics, including energy demand and connection duration;
    \item An ensemble learning approach to improve forecast accuracy and robustness in response to diverse EV user behaviors;
    \item Environment-specific forecasting for the workplace and residential charging settings to enhance demand management precision;
    \item Incorporation of incremental learning and dynamic training updates, ensuring that forecasts remain aligned with the latest trends in EV charging behavior;
    \item Rigorous feature engineering to identify the most influential variables for accurate forecasting;
    \item A dual-model approach for distinct forecasting needs of new and recurring EV users;
    \item An analysis of the impact of data volume on forecast performance to guide optimal data storage and computational efficiency.
\end{enumerate}

This work is structured as follows. Section \ref{method} outlines the methodology and models for session-specific EV charging parameters forecasting. Section \ref{case} introduces the considered case study and describes the considered forecasting features. The results are presented in Section \ref{results}. Finally, Sections \ref{discussion} and \ref{conclusion} are discussions and conclusions including recommendations for future research directions.

\section{Methodology} \label{method}
\begin{figure*}
\centerline{\includegraphics[width=\linewidth]{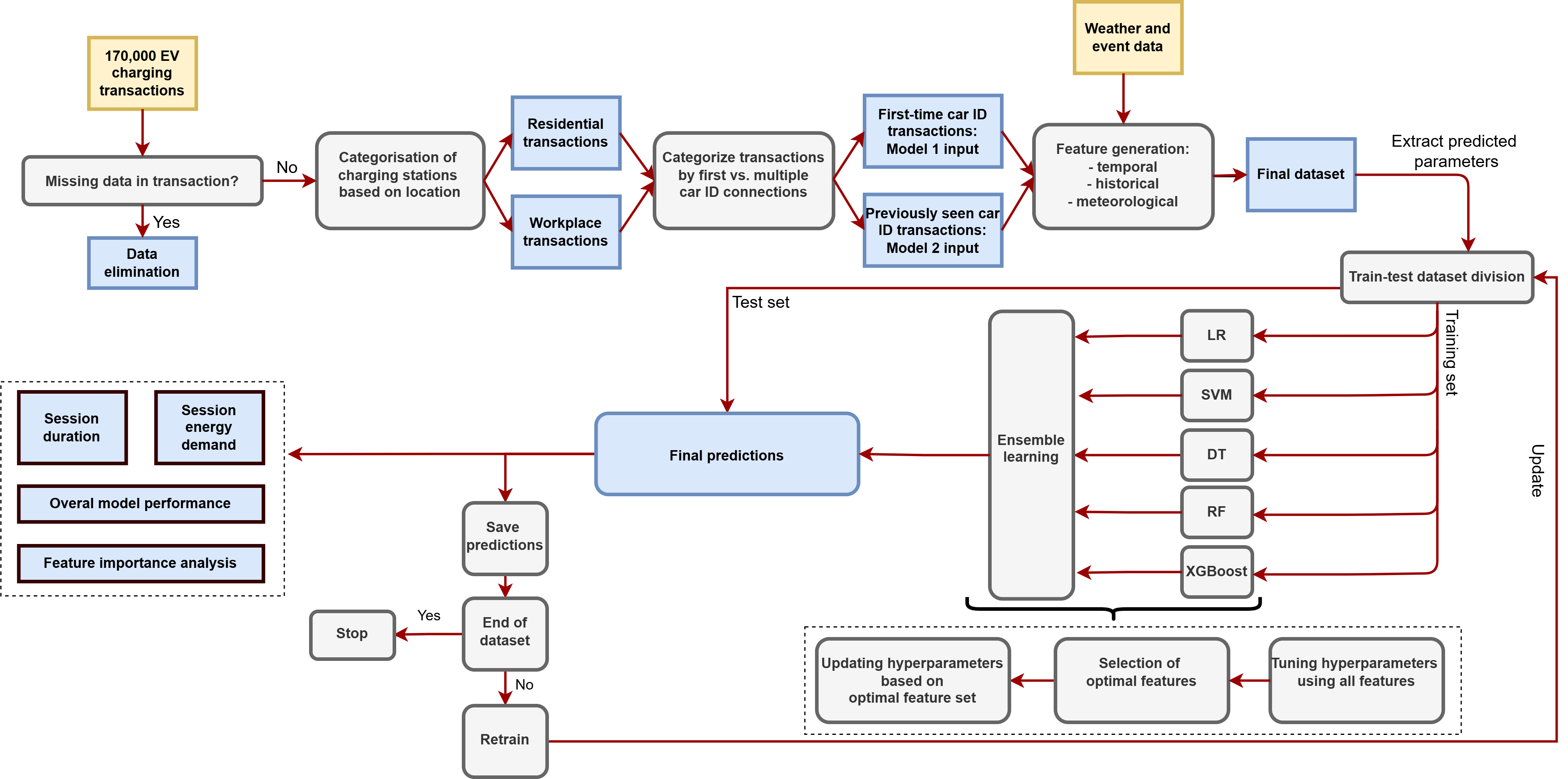}}
\caption{Overview of our proposed framework identifying methodological steps in the analysis.}
\label{fig:method}
\end{figure*}

This section outlines our methodology to identify the optimal forecasting performance, optimal feature sets, feature importance, and the best training dataset size for forecasting EV charging parameters. Fig. \ref{fig:method} presents a summary of the involved methodological steps for forecasting EV charging parameters. In this approach, forecasts for session duration and session energy demand are conducted as separate processes.

We employ a dual-model approach to ensure that optimal forecasting performance is achieved for all EV charging sessions while assuring that no EV charging session from the dataset is excluded. It makes a distinction between EV charging sessions of EVs that have previously been charged at a charging station of the CPO and EVs that are not previously seen in the dataset. In case the ID of an EV has been seen previously in the dataset, it is routed to Model 2, which also considers user-specific features, potentially leading to a better forecasting performance. In the case of new EVs that have never been seen before, Model 1 is used, which operates with a limited set of generic historical features. Through this approach, it is possible to forecast EVs of all types and to achieve the maximum forecasting performance for each EV charging session.

The methodology is implemented on a weekly basis, where all models are trained and tuned, and optimal features are selected for each week. At the start of each iteration, newly logged data from the previous week is added to the training set, and models are retrained to incorporate the most recent patterns. Finally, the forecasting accuracy is determined by combining Model 1 and Model 2 forecasts, which ensures robust metrics across all scenarios.

\subsection{Supervised Machine Learning algorithms}
A supervised learning approach involves learning models using labeled datasets containing the target variable the model is designed to forecast. This section reviews six supervised machine-learning algorithms for forecasting individual EV charging parameters. Linear Regression (LR) captures proportional dependencies between variables, while the non-linear Support Vector Regression (SVR) captures more complex patterns. Decision Tree (DT) and Random Forest (RF) regression models offer robust forecasts by identifying feature interactions. Extreme Gradient Boosting (XGBoost) is applied and known for its high efficiency and accuracy in handling large datasets with complex relationships. Finally, an ensemble learning approach is applied to combine the outputs of these five models to improve forecast accuracy and generalize across different charging patterns.

\subsubsection{Multiple Linear Regression}
An LR model is used to determine how multiple explanatory variables relate to a target variable to forecast it from the values of these explanatory variables. In this case, with \( k \) explanatory variables and \( n \) observations, the model can be represented as follows:

\begin{equation}
y_i = \beta_0 + \sum_{j=1}^{k} \beta_j x_{ij} + \epsilon_i \quad \text{for } i = 1, 2, \dots, n,
\end{equation}

where \( y_i \) is the response variable for observation \( i \), \( \beta_0 \) is the intercept term, \( \beta_j \) are the regression coefficients for each explanatory variable \( x_{ij} \), \( x_{ij} \) represents the value of the \( j \)-th explanatory variable for the \( i \)-th observation, and \( \epsilon_i \) is the error term for observation \( i \), also known as the residual. In this study, we use the Python library \texttt{sklearn.linear\_model.LinearRegression} \cite{buitinck2013api}.

\subsubsection{Support Vector Regression}
SVR is designed to fit data within a specified error margin \( \epsilon \), allowing small deviations while ignoring significant outliers. This approach is controlled by the parameter \( C \), which balances model simplicity against tolerance for errors that exceed the margin. SVR aims to find a function \( y(x) = w \cdot x + b \) that best represents the data by minimizing the objective function below \cite{vapnik2013nature}:

\begin{equation}
\min_{w, b} \left( \frac{1}{2} \|w\|^2 + C \sum_{i=1}^{n} \left( \xi_i + \xi_i^* \right) \right)
\end{equation}

where, \( C \) controls the trade-off between maximizing the margin (keeping \( w \) small) and allowing some errors to exceed \( \epsilon \) through slack variables \( \xi_i \) and \( \xi_i^* \). A higher \( C \) value makes the model less tolerant to errors, aiming for a closer fit to the data but potentially risking over-fitting. The Radial Basis Function kernel is used to capture non-linear relationships by mapping data into a higher-dimensional space, described by:

\begin{equation}
k(x_i, x) = \exp \left(-\gamma \|x_i - x\|^2 \right)
\end{equation}

where \( \gamma \) determines the kernel's spread and influences model complexity. As the training time of SVR is very long, it is not suitable for large datasets. In this study, we use the Python package \texttt{sklearn.svm.SVR} \cite{buitinck2013api}.

\subsubsection{Decision Tree}
DT represents a series of decisions as a tree structure, sequentially splitting the dataset into smaller subsets based on the input features. For each split or decision node, a feature and threshold value are selected. Each leaf node represents a subset of the data and is the endpoint of the tree where no further splits occur. In DT algorithm, forecasts are made by averaging the target values within each leaf node. Based on feature values, the model accumulates a series of decisions until it reaches a leaf, where the forecast is made.

While straightforward to implement, a single decision tree can be prone to over-fitting, especially with complex datasets. Regularization techniques limit tree complexity and avoid over-fitting by limiting the maximum depth of the tree or how many samples are allowed per leaf. In this study, we use the Python package \texttt{sklearn.tree.DecisionTreeRegressor} \cite{cohen2022tree} o develop and optimize the DT regression model.

\subsubsection{Random Forest}
RF regression uses multiple DTs to combine forecasts to create a more accurate and robust model. As RF averages forecasts from a variety of DTs, it reduces the risk of overfitting, making it an effective choice for complex datasets. In an RF model, each tree is built independently on a randomly sampled subset of the data, with each split within a tree chosen based on a random subset of features. Having randomness in the model ensures that each tree captures a variety of patterns and variations within the data, enhancing its generalization abilities. For a given input, the final RF forecast, \( \hat{y} \), is obtained by averaging the forecasts of all individual trees:

\begin{equation}
\hat{y} = \frac{1}{T} \sum_{t=1}^{T} \hat{y}_t
\end{equation}

where \( T \) is the total number of trees in the forest, and \( \hat{y}_t \) is the forecast from the \( t \)-th tree. In this study, we use the Python package \texttt{sklearn.ensemble.RandomForestRegressor} \cite{cohen2022tree}, with parameters such as the number of trees and maximum tree depth optimized to balance model performance and computational efficiency.

\subsubsection{Extreme Gradient Boosting}
With XGBoost, trees are built sequentially instead of independently, as with RF. Each tree corrects the errors of the previous one. This approach, called gradient boosting, minimizes a loss function iteratively to optimize the model. By summing the forecasts of all trees in an ensemble, XGBoost generates a forecast for a given input:

\begin{equation}
\hat{y} = \sum_{t=1}^{T} f_t(x)
\end{equation}

where \( T \) is the total number of trees, \( f_t(x) \) is the forecast from the \( t \)-th tree, and each tree \( f_t \) is trained to minimize the residual errors of the previous trees. In this study, XGBoost regression is implemented using the Python package \texttt{xgboost.XGBRegressor} \cite{cohen2022tree}, where hyperparameters such as learning rate, maximum tree depth, and number of trees are tuned to achieve optimal performance.

\subsubsection{Ensemble Learning}
Ensemble learning leverages the strengths of multiple models to improve forecast accuracy. Ensemble methods enhance model performance by combining forecasts from several learning algorithms. This approach is effective because it reduces reliance on any single model, avoids local optima, and captures complex patterns that might be missed by individual models \cite{dietterich2000ensemble}.

\begin{figure*}
\centerline{\includegraphics[width=0.7\linewidth]{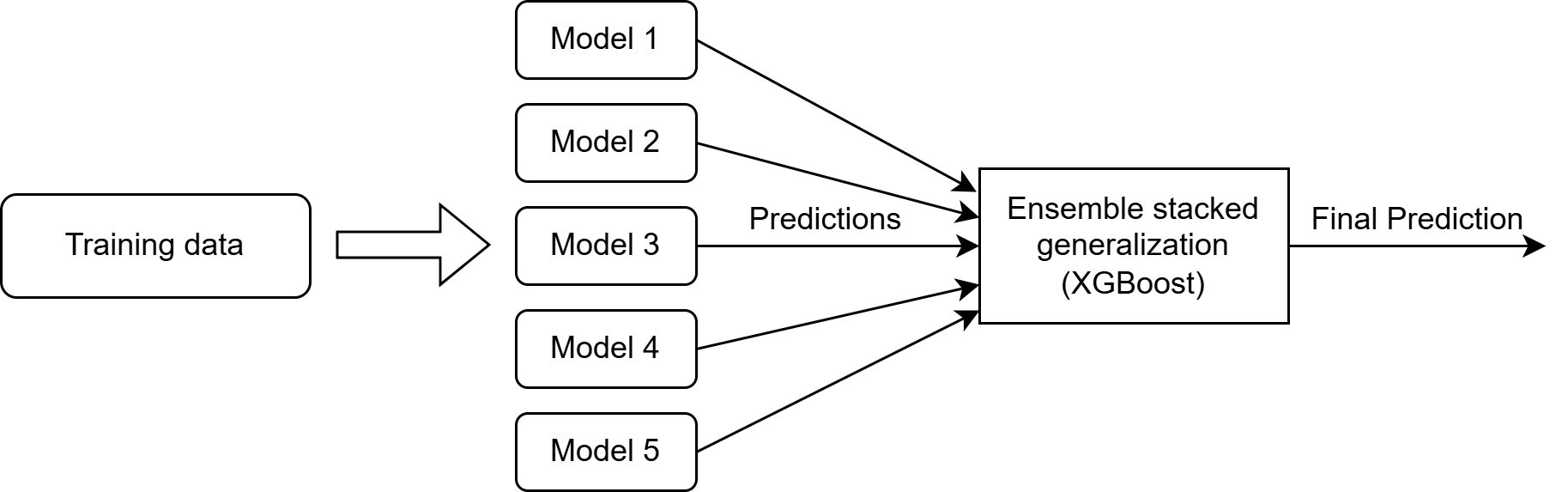}}
\caption{Overview of stacking ensemble learning.}
\label{fig:ensemble}
\end{figure*}

Common ensemble techniques include bagging, boosting, and stacking \cite{divina2018stacking}. With bagging, multiple models are built and then averaged, whereas with boosting, models are weighted based on their performance. Stacking involves combining forecasts from different models using a secondary model to form the final forecast. This study employs a stacking approach, which is well-suited to regression problems. XGBoost is used as the combiner to integrate forecasts from base models. A general scheme of such an approach is shown in Fig. \ref{fig:ensemble}.

\subsection{Hyperparameter tuning}
The performance of various machine learning forecasting models is significantly influenced by their hyperparameters, which must be carefully adjusted to optimize forecasting accuracy. Therefore, hyperparameter tuning is essential for each model and each forecasting target to ensure optimal performance across all considered features. This process involves systematically exploring a wide range of hyperparameter configurations and identifying the combination that yields the best performance for each model. Model performance can be evaluated using metrics such as the coefficient of determination ($R^2$), which offers a normalized score between -1 and 1, enabling consistent performance comparisons across different forecasting targets. By maximizing the $R^2$ score, we ensure each model is well-suited for its specific forecasting task. Hyperparameter tuning was conducted using the sklearn Python package (sklearn.model\_selection.GridSearchCV). The range of considered hyperparameter sets is shown in Table \ref{table:hyper} in Appendix section \ref{app}.

Since the best hyperparameter set may vary depending on the selected features, hyperparameter tuning should also be conducted using the optimal feature subset explained in section \ref{feature}. This iterative tuning process ensures that each model’s performance is tailored not only to the forecasting target but also to the most informative feature set, minimizing overfitting and improving the robustness of forecasting outcomes.

\subsection{Feature selection} \label{feature}
In certain cases, specific features may negatively impact a forecasting model’s performance, often due to factors like multicollinearity \cite{raschka2019python}. Therefore, identifying the optimal feature set for each forecasting model is essential. A widely used approach for feature selection is sequential (floating) feature selection, which can be implemented in either a forward or backward manner, as described in \cite{biom10030454}. Sequential (floating) feature selection methods are iterative approaches that dynamically adjust the feature set to identify the most informative subset. In forward sequential selection, features are progressively added, one at a time, based on their ability to improve model performance, while in backward sequential selection, features are removed iteratively if they no longer contribute meaningfully to model accuracy. The feature selection was performed using sequential floating backward feature selection, using the MLxtend Python package \cite{raschka2018mlxtend}.

To reduce the likelihood of model overfitting, these methods should be applied to the training dataset using k-fold cross-validation. After determining potential feature sets, model performance can be evaluated by applying the trained model to the test dataset, allowing the forecasting accuracy with different feature configurations to be compared. Repeating these steps across multiple iterations increases the robustness and reliability of the results. The feature subset with the highest average performance on the test dataset should be selected as the optimal set for the forecasting model. Additionally, for tree-based models, feature importance was evaluated using the model’s built-in attribute, which quantifies each feature’s relative contribution to the forecasting process.

\subsection{Incremental learning}
With an incremental learning approach, the model is continually updated with newly logged data while retaining a representative subset of past examples \cite{luo2020appraisal}. This allows the model to balance learning new patterns while mitigating catastrophic forgetting, a phenomenon where previously learned information is lost when adapting to new data.. Retraining is accomplished by storing and replaying a small selection of key past samples alongside new data rather than retraining on the entire historical dataset. In this framework, training is conducted iteratively on a weekly basis, where the models are retrained using newly logged data from the previous week. This iterative process ensures the model remains responsive to recent changes while efficiently managing memory constraints, making it well-suited for dynamic and evolving datasets such as EV charging patterns.

\subsection{Dynamic temporal data splitting}
A data splitting strategy was used in this study to accommodate the temporal nature of the dataset and ensure that the models were trained and evaluated robustly. Splitting is a dynamic process in which the training and validation datasets continually grow. Initially, the model is trained using one year of historical data, with the first 80\% for training and the last 20\% for validation. The model is then tested on data from the following week. In subsequent iterations, the training and validation datasets are extended by incorporating the previously tested data into the training set. For example, in the second iteration, the model is trained and validated using one year and one week of data, then tested using the following week. As each step of the process is completed, the training and validation sets dynamically grow by one week. Therefore, this approach does not maintain a constant proportion of training, validation, and test datasets. Instead, the data split percentages change gradually as more data is added to the training and validation sets, enhancing the model's learning capabilities.

\subsection{Forecast Performance evaluation}
To evaluate the predictive accuracy of the models, several performance metrics are applied, as outlined in \cite{botchkarev2019new}. In this study, we utilize three metrics commonly employed in related work. Lower Root Mean Square Error (RMSE) and Mean Absolute Error (MAE) scores indicate high forecast accuracy, while the Coefficient of Determination (\( R^2 \)) value, ranging from -1 to 1, assesses the model fit, with 1 representing perfect forecasts and higher values indicating better performance.

This study evaluates forecasting performance across all EV types using the combined outputs of Model 1 and Model 2. As a result, the final performance metrics accurately reflect the forecasting capabilities of both models, integrating results for previously seen and new EVs into one comprehensive analysis.

\section{The case study} \label{case}
\subsection{Data collection}
This study analyzed historical data from EV charging sessions collected at charging stations managed by the CPO We Drive Solar \cite{wedrive}. These stations, located on public streets in Utrecht, the Netherlands, were accessible to all EV users without the requirement for a subscription. Users were charged a fixed rate per kWh, and the stations supported a maximum power output of 22 kW, thereby excluding fast-charging capabilities.

The dataset includes detailed records of each charging session, capturing information such as arrival time, departure time, energy demand, EV ID, charging station ID, station location, and the maximum charging power per session. The dataset covers the period from January 1, 2022, to December 31, 2023, and includes approximately 170,000 charging sessions. Of these, 160,000 sessions are from residential areas, while 10,000 sessions are from workplace charging stations. The workplace stations in the dataset specifically represent the parking lot of a banking company, chosen to provide insights into typical charging patterns at corporate facilities and office environments \cite{van2021impact}. Charging sessions were categorized as either residential-based or workplace-based according to a method described in \cite{panda2024quantifying}, which uses criteria like location, accessibility, and usage patterns. The dataset exclusively represents charging sessions from stations managed by a single operator located in various places with differing access restrictions. Based on accessibility tags and location type, stations were classified as either residential or workplace chargers.

In addition, we identified and removed outliers in energy demand, as some recorded charging demands were not reasonable given the duration of the respective sessions. The connection duration was required to be sufficient to meet the energy demand, considering the maximum charging power of the station. Furthermore, sessions with energy demands of less than 1 kWh or connection durations shorter than 15 minutes were excluded from the dataset to eliminate unrealistic or insignificant charging sessions.

\subsection{Understanding user behavior}
\begin{figure*}[!b]
\centering
\begin{minipage}[t]{0.4\textwidth}
    \includegraphics[width=\linewidth]{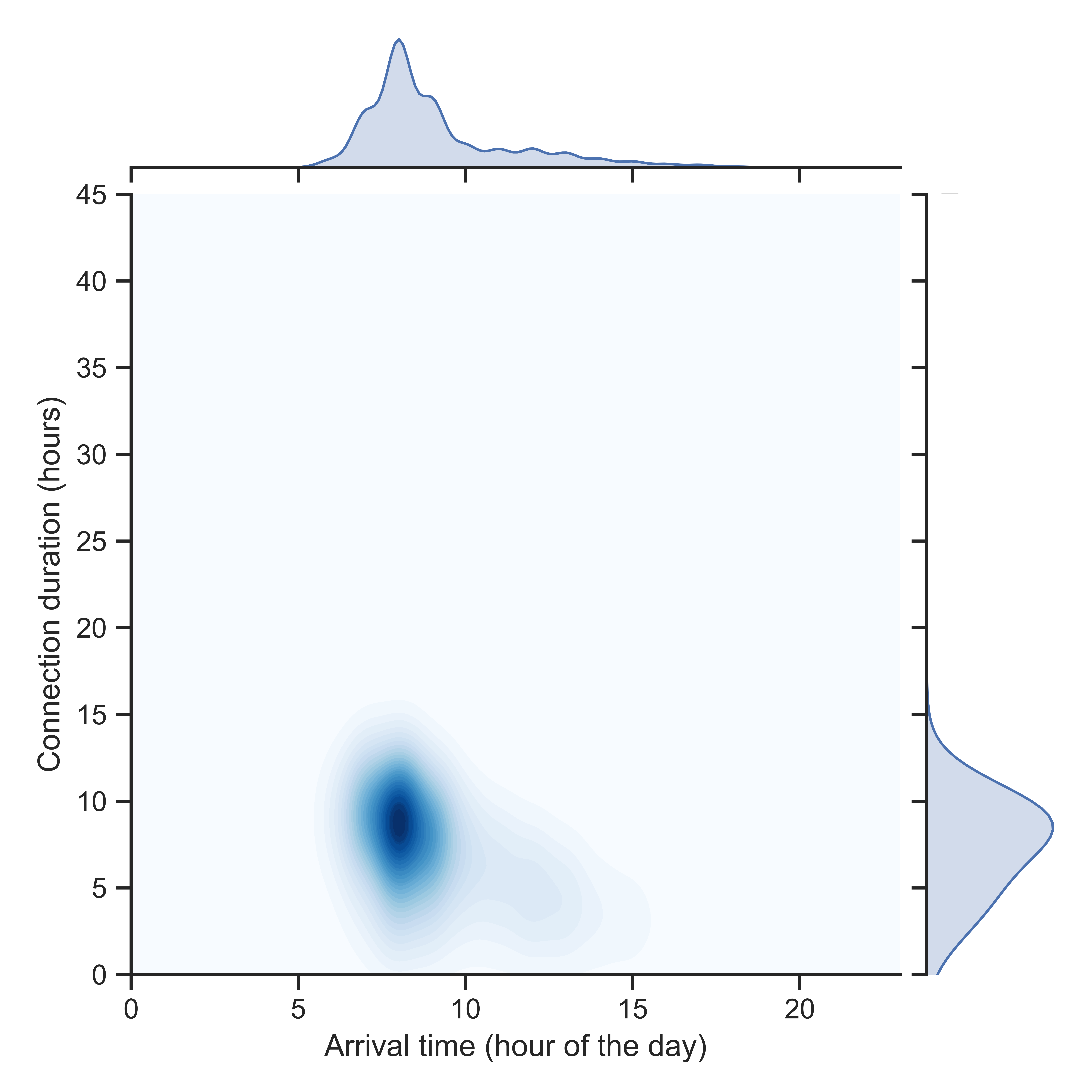}
\end{minipage}\hspace{0.05\textwidth}
\begin{minipage}[t]{0.4\textwidth}
    \includegraphics[width=\linewidth]{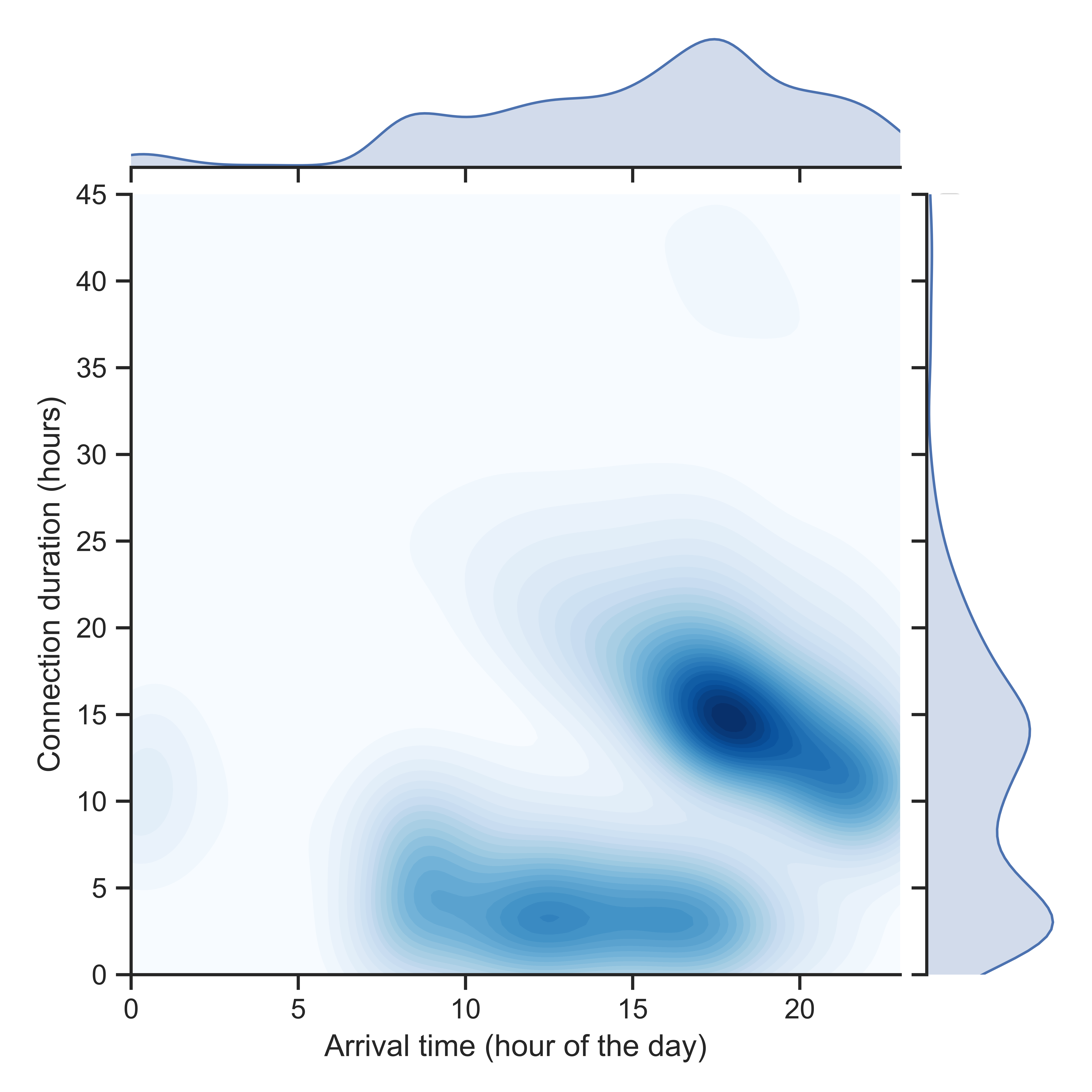}
\end{minipage}
\caption{EV arrival hours and connection durations for workplace (left) and residential (right) charging stations.}
\label{fig: parameters distribution}
\end{figure*}

Fig. \ref{fig: parameters distribution} presents the distribution of EV arrival times and connection durations for workplace charging stations (left) and residential charging stations (right). The data shows a clear pattern in workplace locations, where EVs typically arrive between 7-9 AM, corresponding with standard commuting hours. Vehicles at workplace stations tend to stay connected for extended periods, with a peak connection duration of approximately 8-10 hours, reflecting a typical workday length. Residential charging stations display a different usage pattern, with two distinct peaks in connection duration: a shorter duration, around 3 hours, and a longer duration, around 16 hours. This dual peak suggests varied charging behaviors, potentially due to short stops and overnight charging. The differences between workplace and residential charging behaviors highlight the impact of location on charging patterns, with workplace stations supporting day-long charging during business hours and residential stations accommodating both brief evening charges and overnight stays.

\begin{figure*}[!b]
\centering
\begin{minipage}[t]{0.4\textwidth}
    \includegraphics[width=\linewidth]{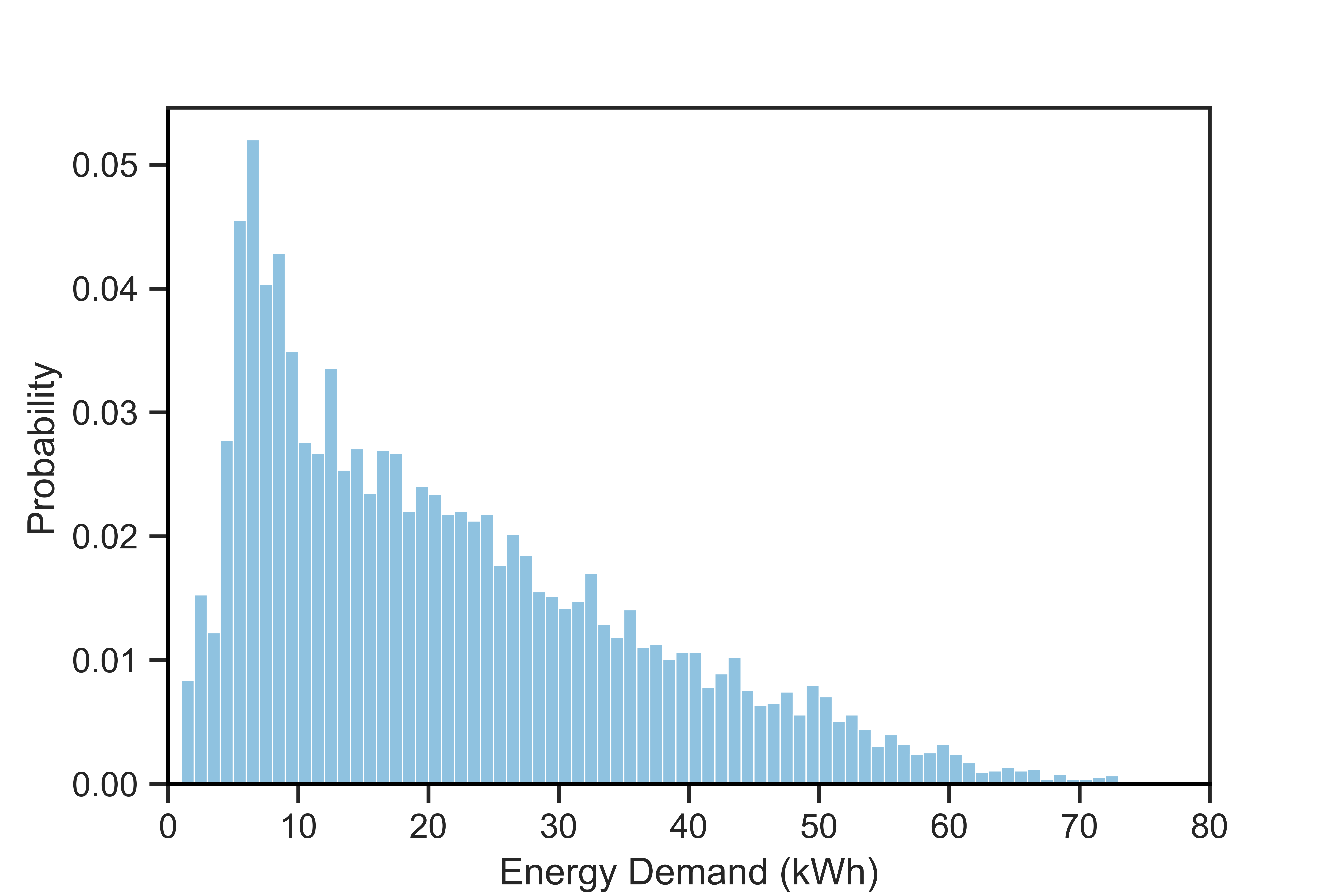}
\end{minipage}\hspace{0.05\textwidth}
\begin{minipage}[t]{0.4\textwidth}
    \includegraphics[width=\linewidth]{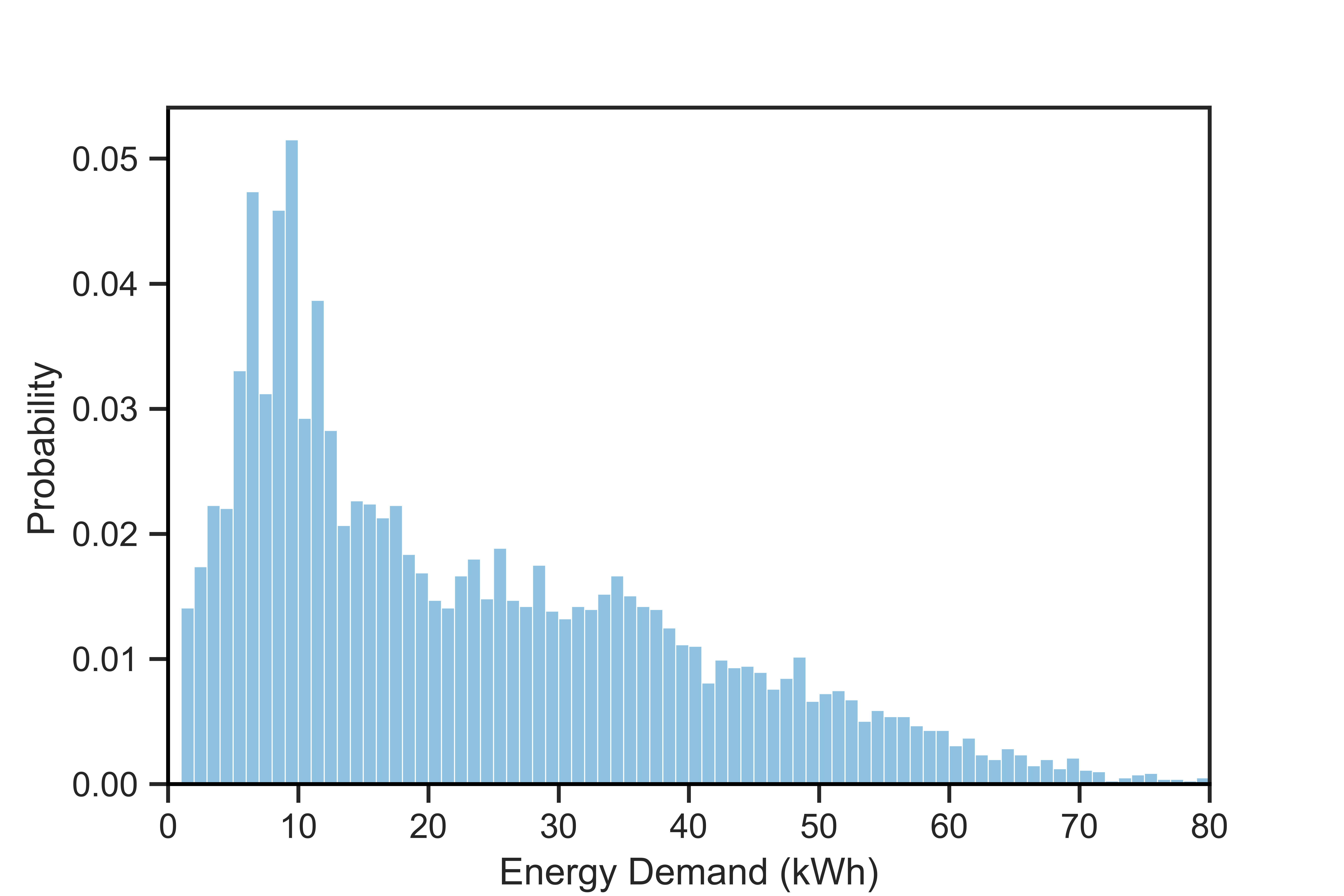}
\end{minipage}
\caption{Energy demand distribution at workplace (left) and residential (right) charging stations.}
\label{fig: energy demand}
\end{figure*}

The energy demand distribution for EV charging sessions at workplace and residential stations reveals their usage patterns, as shown in Fig. \ref{fig: energy demand}. At workplace stations (left), the energy demand peaks between 7-10 kWh, reflecting moderate energy requirements. In contrast, residential charging stations (right) display a broader distribution of energy demand, with a peak of around 6-13 kWh. This higher and wider range indicates more variability in charging behaviors, as residential charging accommodates both shorter, partial charges and longer, full charges, likely due to overnight stays and varying state-of-charge levels upon arrival.

\begin{figure*}[!b]
\centering
\begin{minipage}[t]{0.4\textwidth}
    \includegraphics[width=\linewidth]{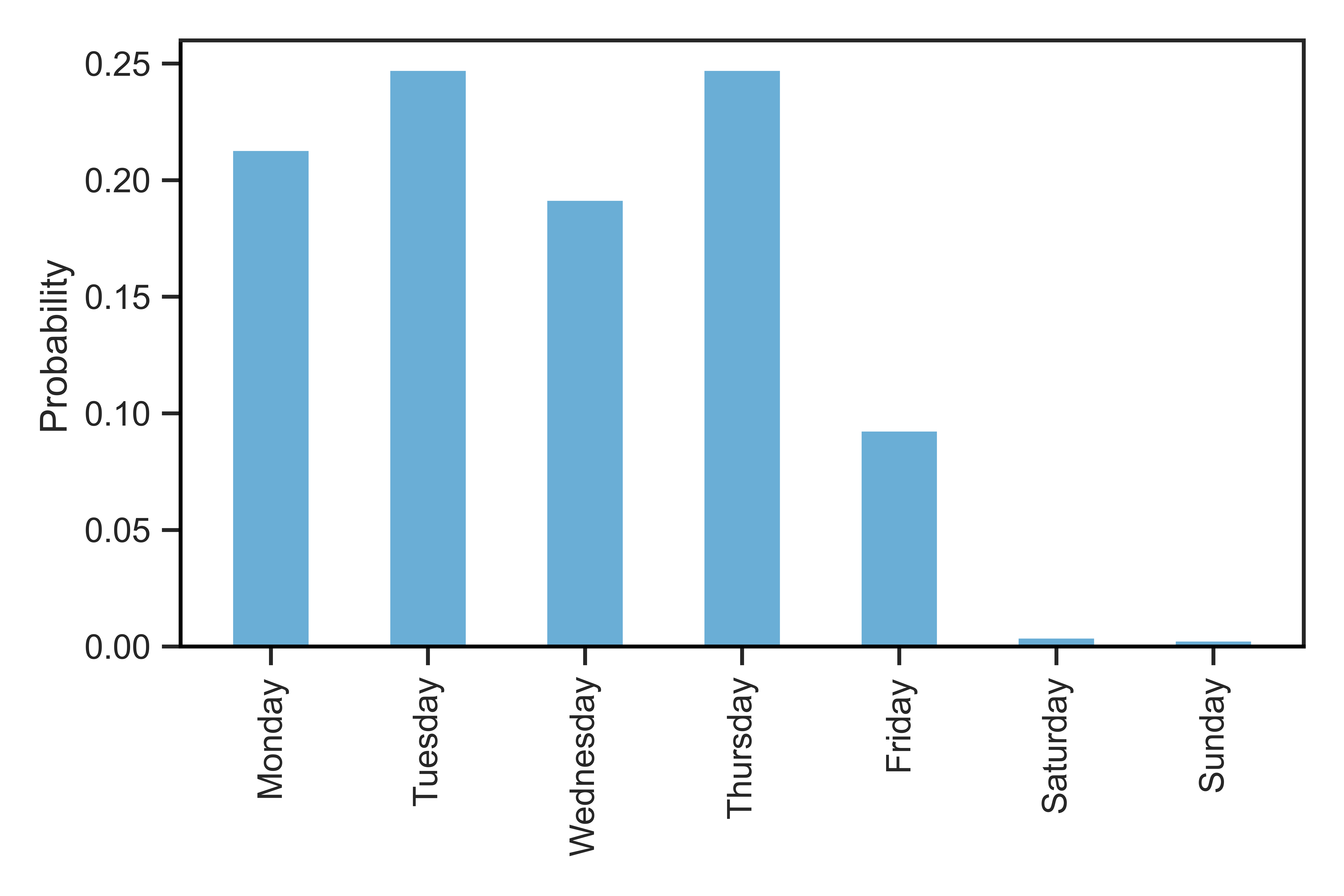}
\end{minipage}\hspace{0.05\textwidth}
\begin{minipage}[t]{0.4\textwidth}
    \includegraphics[width=\linewidth]{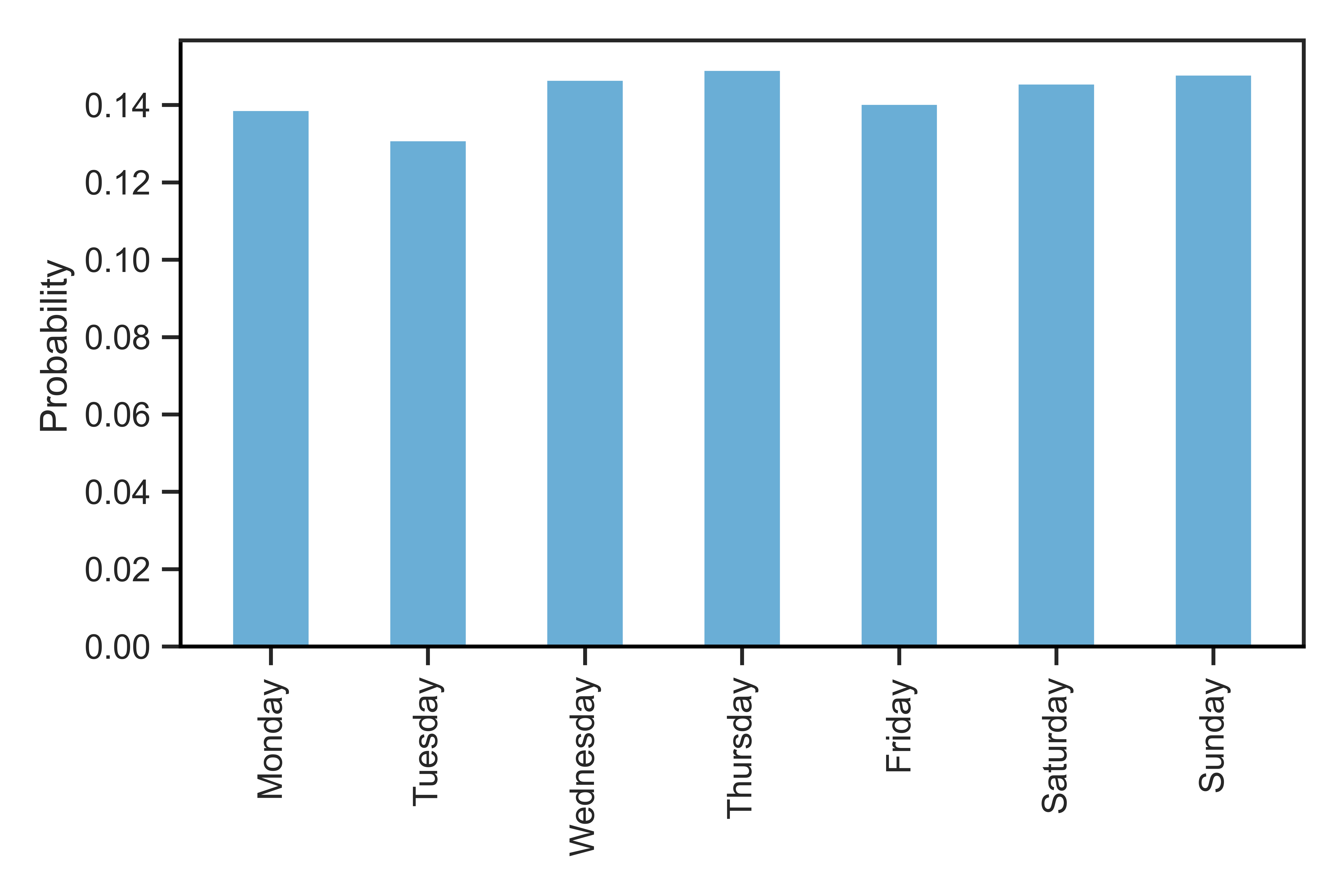}
\end{minipage}
\caption{Daily distribution of EV charging sessions at workplace (left) and residential (right) stations.}
\label{fig: day of week}
\end{figure*}

The distribution of EV charging sessions per day of the week shows distinct patterns between workplace and residential charging stations, as shown in Fig. \ref{fig: day of week}. Workplace stations experience high session volumes from Monday to Thursday, with a noticeable decline on Friday and very limited activity over the weekend. This trend aligns with typical work schedules, where EV users rely on workplace charging during standard weekdays. On the other hand, residential charging stations maintain a relatively steady session rate throughout the entire week, with similar levels of activity across all days. This consistent usage pattern suggests that residential charging is less influenced by the weekday-weekend divide. These differences emphasize how location affects charging behavior, with workplace stations reflecting structured weekday demand and residential stations supporting continuous use.

\subsection{Features}
Three main categories of features are utilized to forecast EV charging parameters. The first category includes temporal features, such as the hour of the day and day-type (i.e., weekday, weekend, school holiday, or public holiday). The second category leverages historical data as features, incorporating past values of the forecasted parameter, like the average value per station ID and car ID. The third category involves day-ahead weather forecasts, including temperature, precipitation volume, and average wind speed, since weather conditions can influence transport mode choices. Table \ref{tab:feature_categories} provides an overview of all 15 features considered. Historical features related to each car and station ID (features 8, 9, 11, and 12) are considered only in Model 2, where the parameters of previously seen cars are forecasted.

\begin{table}[!b]
\centering
\caption{Overview of the considered features for the forecasting of EV charging parameters. Features in bold are considered only in Model 2.}
\label{tab:feature_categories}
\begin{tabular}{>{\raggedright}p{2cm}>{\raggedright}p{0.5cm}>{\raggedright}p{7.5cm}>{\raggedright}p{1.5cm}>{\raggedright\arraybackslash}p{1.5cm}}
\hline
\textbf{Category} & \textbf{No.} & \textbf{Explanation} & \textbf{Short notation} & \textbf{Source} \\
\hline
\multirow{4}{3cm}{Temporal} & 1 & Hour of the day & HOUR & - \\
 & 2 & Month of the year & MONTH & - \\
 & 3 & Season of the year & SEASON & - \\
 & 4 & Day of the week & T:WD & - \\
 & 5 & Whether the considered day is a national holiday & T:NH & - \\
 & 6 & Whether the considered day is a school holiday & T:SH & - \\
 & 7 & Whether the considered day is Weekend & T:W & - \\
\hline
\multirow{4}{3cm}{Historical and descriptive} & 8 &  \textbf{Average of parameter per car ID} & H:AVC & - \\
 & 9 &  \textbf{Average of parameter per station ID} & H:AVS & - \\
 & 10 & Average of parameter per hour & H:AVH & - \\
 & 11 &  \textbf{Maximum of parameter per CarID} & H:MAX & - \\
 & 12 &  \textbf{Minimum of parameter per CarID} & H:MIN & - \\
\hline
\multirow{4}{3cm}{Meteorological} & 13 & Hourly temperature & M:T & KNMI\cite{KNMI2024} \\
 & 14 & Hourly wind speed & M:WS & KNMI\cite{KNMI2024} \\
 & 15 & Hourly precipitation volume & M:PV & KNMI\cite{KNMI2024} \\
\hline
\end{tabular}
\end{table}

\section{Results} \label{results}
\subsection{Overall model performance}
\begin{figure*}[t]
\centering
\begin{minipage}[t]{0.45\textwidth}
    \includegraphics[width=\linewidth]{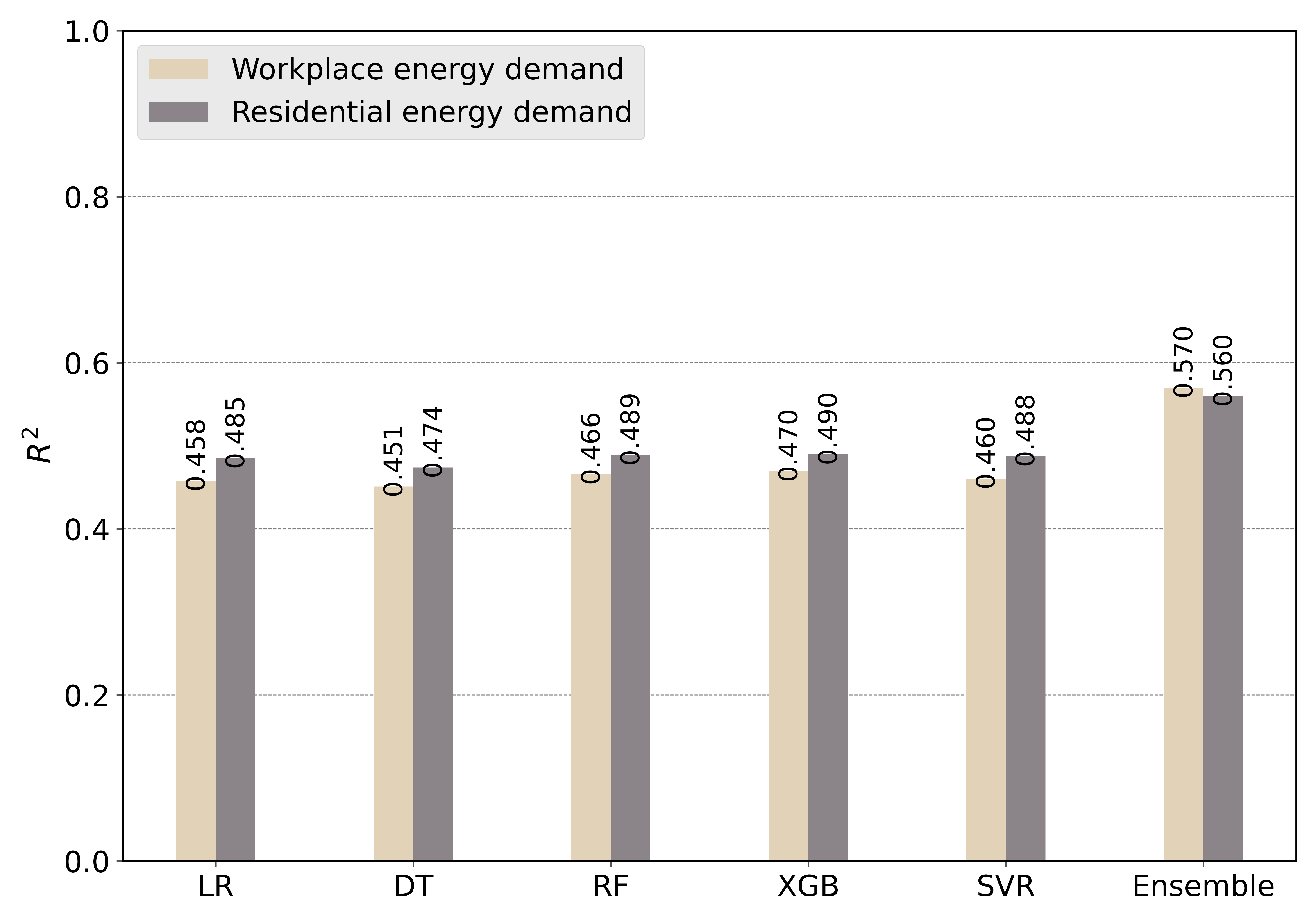}
\end{minipage}\hspace{0.05\textwidth}
\begin{minipage}[t]{0.45\textwidth}
    \includegraphics[width=\linewidth]{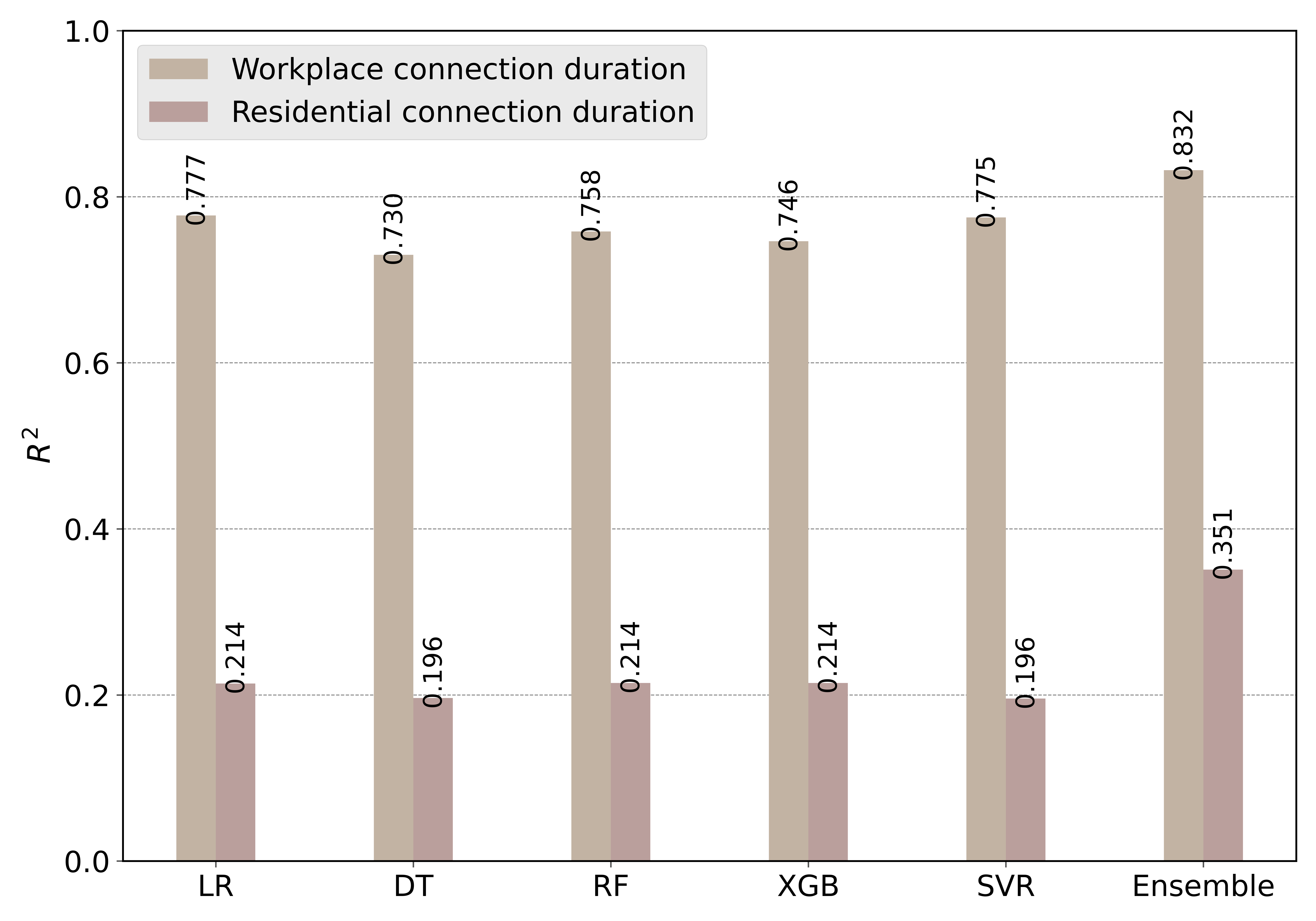}
\end{minipage}
\caption{Average forecasting performance (expressed as the $R^2$ value) for the different considered forecasting models for the different EV charging parameters and locations.}
\label{fig: R2 }
\end{figure*}

Fig. \ref{fig: R2 } shows $R^2$ values for residential as well as workplace session datasets for predicting energy demand and connection duration. The RMSE and MAE metrics for forecasts generated by the five trained models are summarized in Tables \ref{tab: performance metrics energy} and \ref{tab: performance metrics duration} as well.

For energy demand forecast across both workplace and residential locations, all models exhibit mostly similar performance metrics, with RMSE, MAE, and $R^2$ scores showing minimal variation among them. The XGBoost and RF models achieve the lowest RMSE and MAE values, indicating a slightly higher capability to capture complex patterns in energy demand. In addition, a comparable trend is observed for the connection duration forecast. In workplace settings, SVR and RF have the lowest RMSE and MAE values, suggesting that they are more effective at capturing the variability in connection durations. Across residential locations, the models perform similarly, with only slight variations in accuracy, suggesting that simpler models, including LR, can achieve competitive performance. The results indicate that, while some models perform better than others in terms of energy demand and connection duration, overall model performance is relatively consistent.

Results show that while forecasting energy demand at workplace and residential charging locations is relatively comparable, significant differences emerge in forecasting connection duration. Residential locations have much higher errors when forecasting connection duration than workplace locations. This disparity can be attributed to greater variability in residential charging behaviors, driven by diverse user schedules, unpredictable arrival and departure times, and varied charging needs. In contrast, workplace charging follows more structured and predictable patterns, often aligned with standard work hours, leading to improved forecasting accuracy. For workplace charging, the inter-quartile range (IQR) is 4 hours (Q1 = 5.5 hours, Q3 = 9.5 hours), indicating a tight clustering of durations around typical work shifts. On the other hand, residential charging exhibits a much wider IQR of 12 hours (Q1 = 4.25 hours, Q3 = 16.25 hours), reflecting more significant variability in charging behavior.

\begin{table*}[t]
\centering
\caption{Average forecasting performance (RMSE and MAE) of various models for energy demand forecast across different EV charging locations.} 
\label{tab: performance metrics energy}
\begin{tabular}{|*{13}{c|}}
\hline
\multicolumn{1}{|c|}{} & \multicolumn{6}{|c|}{Workplace} & \multicolumn{6}{c|}{Residential} \\ \hline
\multicolumn{1}{|c}{} & \multicolumn{3}{|c}{mean(kWh)} & \multicolumn{3}{|c}{21.13} & \multicolumn{3}{|c}{mean(kWh)} & \multicolumn{3}{|c|}{24.03} \\ \hline
\multicolumn{1}{|c}{} & \multicolumn{3}{|c}{std.(kWh)} & \multicolumn{3}{|c}{14.49} & \multicolumn{3}{|c}{std.(kWh)} & \multicolumn{3}{|c|}{16.97} \\ \hline
Algorithm & LR & SVR & DT & RF & XGB & Ensemble & LR & SVR & DT & RF & XGB & Ensemble \\ \hline
RMSE(kWh) & 9.71 & 9.69& 9.78& 9.63& 9.60& 8.56& 11.15& 11.12& 11.28& 11.11& 11.09&10.11\\ \hline
MAE(kWh) & 6.69 & 6.37& 6.60& 6.47& 6.48&5.96& 7.98& 7.71& 7.95& 7.79& 7.80&7.24\\ \hline
\end{tabular}
\end{table*}

\begin{table*}[t]
\centering
\caption{Average forecasting performance (RMSE and MAE) of various models for connection duration forecast across different EV charging locations.} 
\label{tab: performance metrics duration}
\begin{tabular}{|*{13}{c|}}
\hline
\multicolumn{1}{|c|}{} & \multicolumn{6}{|c|}{Workplace} & \multicolumn{6}{c|}{Residential} \\ \hline
\multicolumn{1}{|c}{} & \multicolumn{3}{|c}{mean(hour)} & \multicolumn{3}{|c}{8.87} & \multicolumn{3}{|c}{mean(hour)} & \multicolumn{3}{|c|}{12.11} \\ \hline
\multicolumn{1}{|c}{} & \multicolumn{3}{|c}{std.(hour)} & \multicolumn{3}{|c}{9.99} & \multicolumn{3}{|c}{std.(hour)} & \multicolumn{3}{|c|}{10.27} \\ \hline
Algorithm & LR & SVR & DT & RF & XGB& Ensemble & LR & SVR & DT & RF & XGB& Ensemble \\ \hline
RMSE(hour) & 4.13 & 4.16& 4.61& 4.83& 4.45&3.46& 7.86& 7.96& 7.96& 7.85& 7.87&6.61\\ \hline
MAE(hour) & 1.73 & 1.54& 1.73& 1.56& 1.63&1.08& 4.81& 4.65& 4.83& 4.68& 4.71&3.91\\ \hline
\end{tabular}
\end{table*}

Each of the five models predicts EV charging parameters based on different modeling techniques and sensitivities to underlying feature relationships, such as linearity, non-linearity, and interactions among variables. The percentage of times each model achieves the closest forecast to the actual value is different, reflecting their distinct strengths in capturing specific patterns in the data. It is interesting to note that SVR records the highest percentage of individual forecasts that are closest to the actual value, while XGBoost has the lowest. However, XGBoost shows the best overall performance across all datasets. This difference highlights an important distinction: even though SVR often achieves higher accuracy, its occasional extreme outlier forecasts negatively affect its overall performance. Contrary to XGBoost, which achieves the closest forecast less frequently, it maintains consistent accuracy with fewer severe outliers, leading to better overall performance.

As a result of this analysis, ensemble learning enhances accuracy by utilizing the unique patterns captured by each model. In our approach, we utilize ensemble learning by adding the forecasts of each model to a meta-model as shown in Fig. \ref{fig:ensemble}, which learns to optimize the final forecast by accounting for the strengths and weaknesses of each base model. The original features are included alongside the model forecasts to provide the meta-model with comprehensive information, allowing it to refine forecasts further. An ensemble learning model's performance metrics across all forecasting parameters and locations are also presented in Fig. \ref{fig: R2 }, Tables \ref{tab: performance metrics energy} and \ref{tab: performance metrics duration}.

Compared with the best individual models, the ensemble model achieves higher $R^2$ values. Based on the best models, energy demand forecasts for workplaces and residential locations were 0.50 and 0.52, respectively, which increased to 0.57 and 0.56, respectively, with the ensemble model. Similarly, the best individual models obtained $R^2$ values of 0.80 for workplace and 0.24 for residential, while the ensemble model achieved improved values of 0.83 for workplace and 0.35 for residential. Both energy demand and connection duration forecasts improve as a result of ensemble learning by capturing a broader range of patterns. Such improvements, particularly in connection duration at residential locations, given the lower baseline accuracy in this context, are notable and represent a meaningful step forward in capturing the complex charging patterns that individual models struggle with. In forecasting studies, an $R^2$ improvement of this magnitude is considered substantial, especially in domains where high variability and external influencing factors make forecasting inherently difficult.

In our dual-model framework, we implemented two models based on whether the Car ID had been previously seen at the charging station. When a Car ID was recognized, historical data related to that vehicle was used as input features (Model 2). Otherwise, the forecast relied on a secondary model (Model 1), which lacked car- and station-specific historical data. The results indicate that Model 2 significantly outperforms Model 1, and the overall presented results reflect a mixture of both models' usage across EV arrivals. However, Model 1 exhibits extremely low predictive performance, with $R^2$ values ranging between 0.02 and 0.07 across all forecasting models, making it largely ineffective for accurate forecasts. This lower performance negatively impacts the overall accuracy of our framework, emphasizing the critical role of historical data from individual cars and stations in improving forecasting reliability and ensuring effective planning.

\begin{figure*}
\centering
\begin{minipage}[t]{0.7\textwidth}
    \centering
    \includegraphics[width=\linewidth]{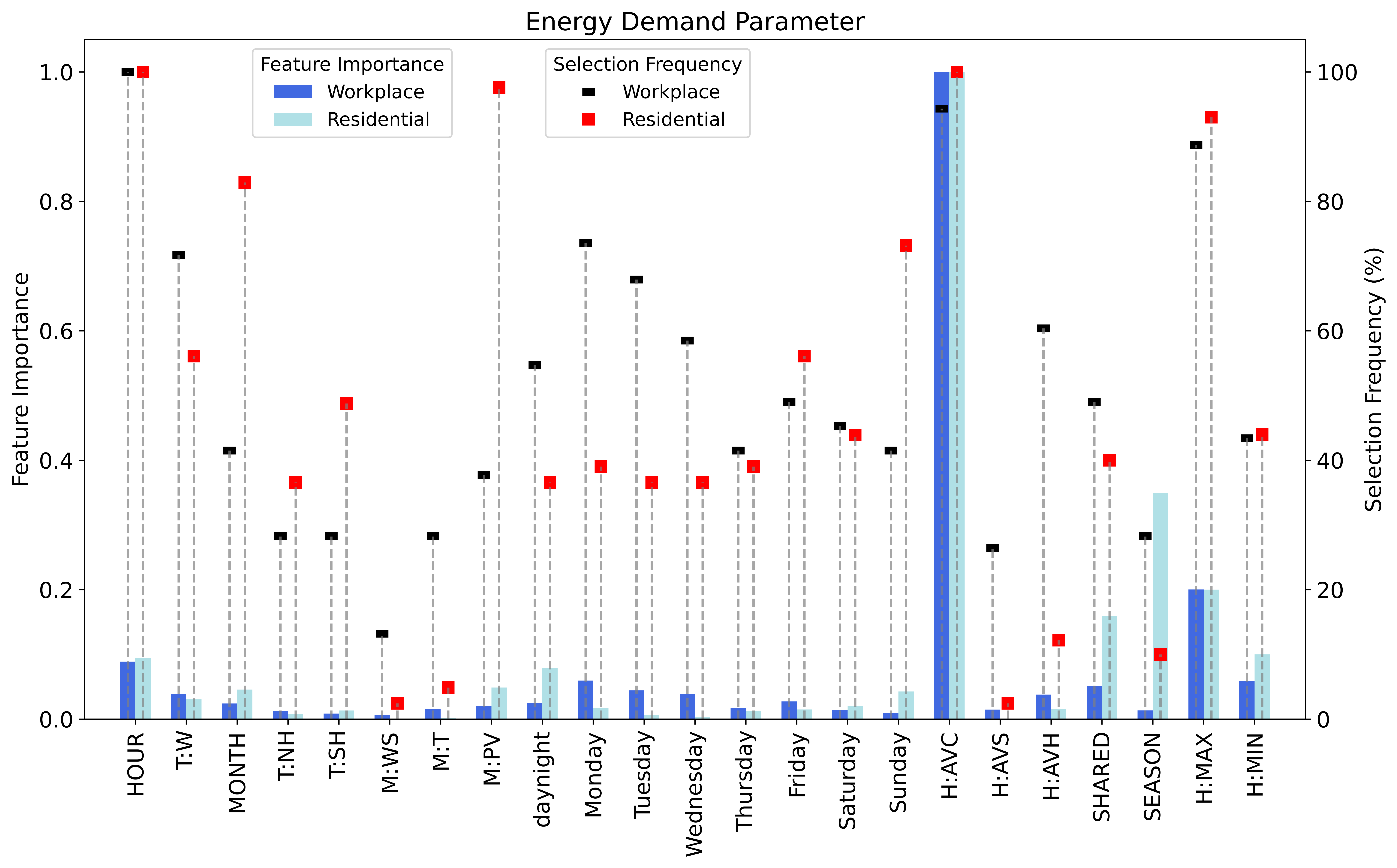}
\end{minipage}
\vspace{0.5cm}
\begin{minipage}[t]{0.7\textwidth}
    \centering
    \includegraphics[width=\linewidth]{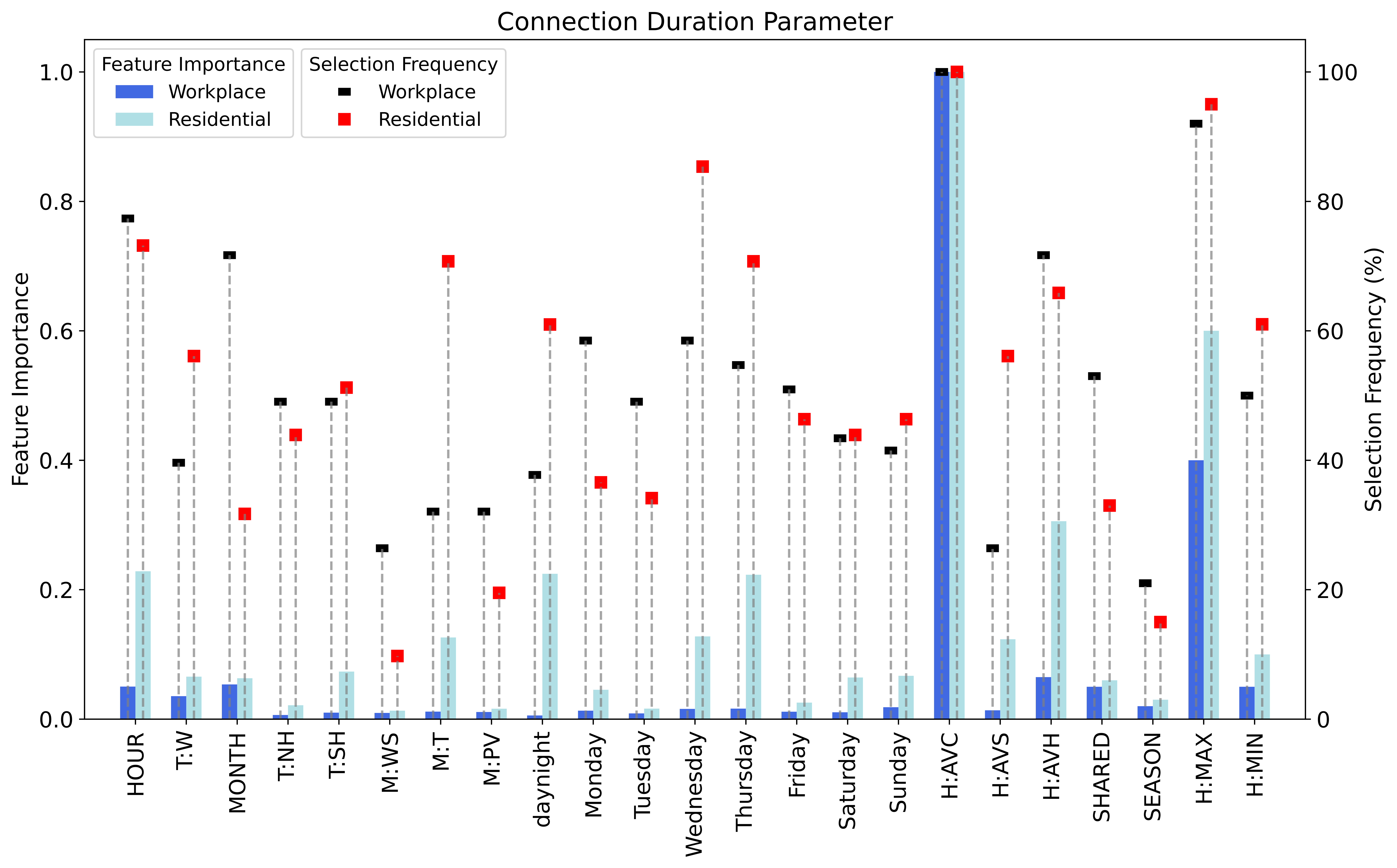}
\end{minipage}
\caption{Feature importance and selection frequency for the base models.}
\label{fig: importance1}
\end{figure*}

\begin{figure*}
\centering
\begin{minipage}[t]{0.49\textwidth}
    \includegraphics[width=\linewidth]{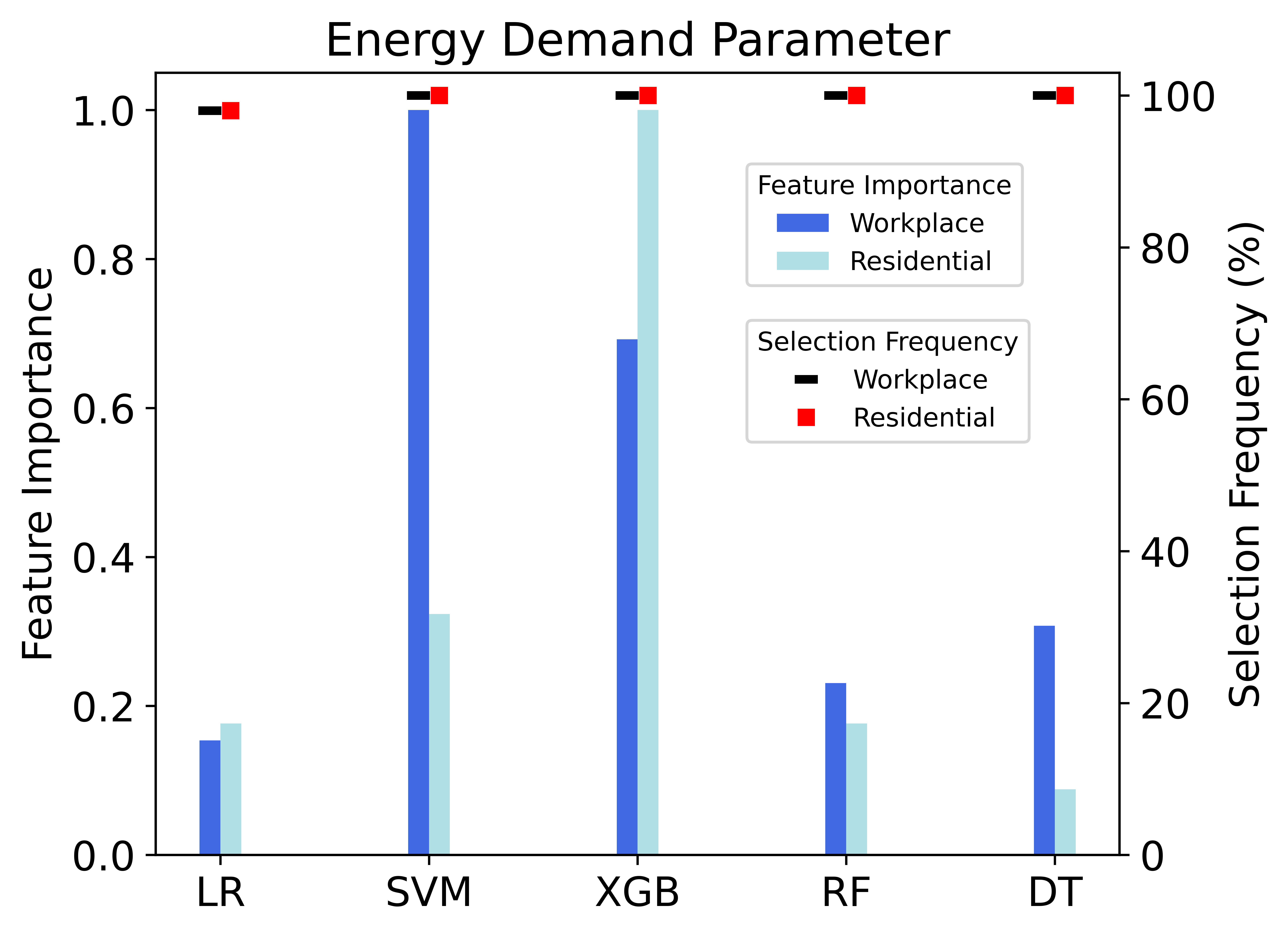}
\end{minipage}\hspace{0.001\textwidth}
\begin{minipage}[t]{0.49\textwidth}
    \includegraphics[width=\linewidth]{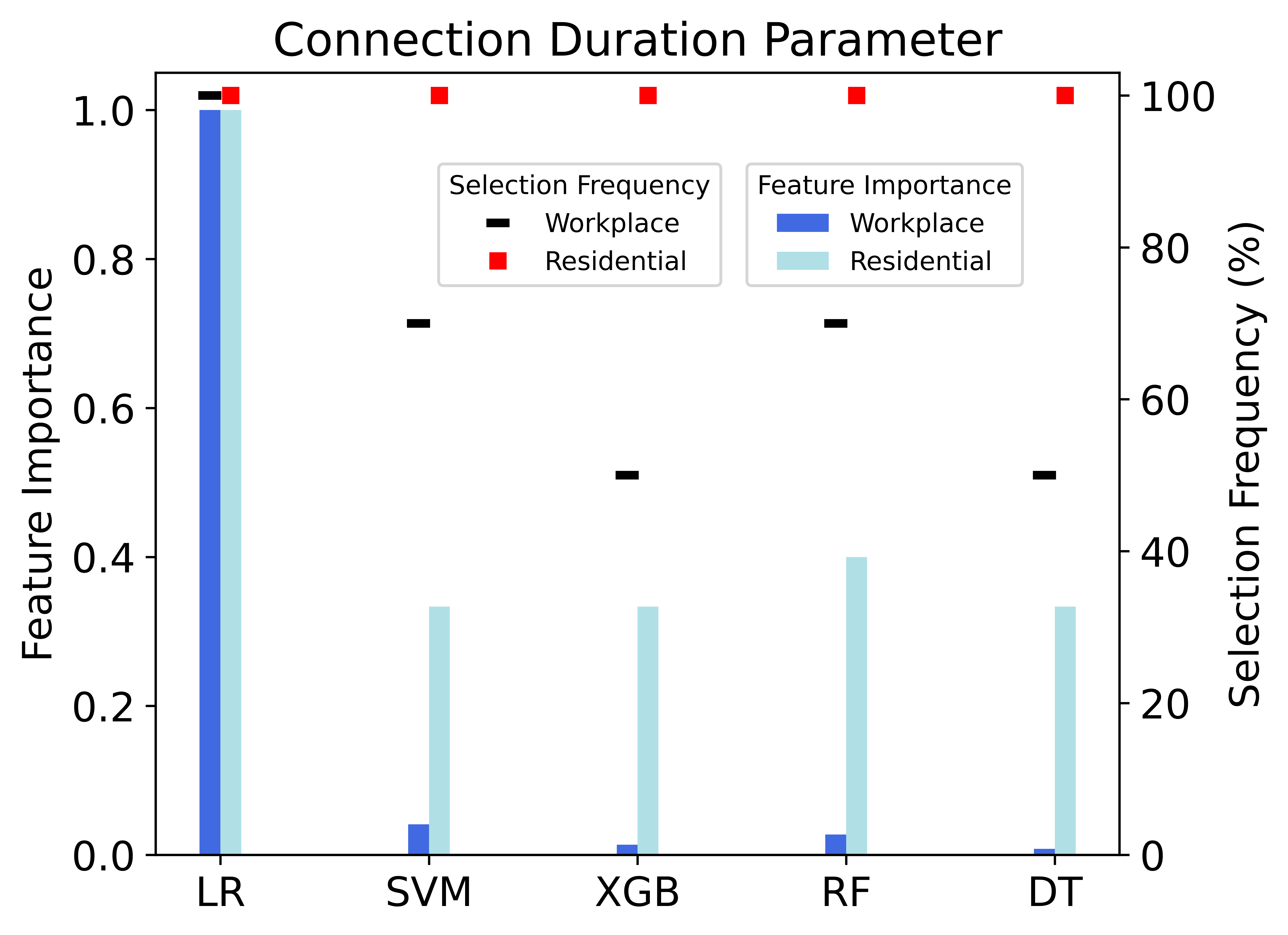}
\end{minipage}
\caption{Feature importance and selection frequency of base models used as features in the ensemble model.}
\label{fig: importance2}
\end{figure*}

\subsection{Feature selection and importance}
The optimal feature set for each forecasting model and each week-ahead forecast was selected and used in model training and forecasting. The unique feature set was determined on a weekly basis over the study period of one year. For each week, the model was retrained, and during each weekly training iteration, the optimal feature sets were dynamically selected based on their contribution to five base models and ensemble learning. Fig. \ref{fig: importance1} shows the average selection frequency and importance of each feature, aggregated across all weekly training iterations for all base models. The left Y-axis represents normalized feature importance, highlighting which features are most critical on average across all weekly training sessions for the week-ahead forecasts. The right Y-axis indicates how frequently features were selected among all iterations.

Certain historical and temporal features are consistently selected as the most influential for both forecasting parameters across most iterations. Specifically, the hour of the day, maximum of the parameter per Car ID, and average of the parameter per Car ID show the highest selection frequency, highlighting their significance in the forecast model. Additionally, for connection duration in residential areas, weekdays are found to be a highly selected factor with an average importance. In contrast, meteorological factors play a minor role, while temporal and historical features—particularly those related to Car IDs—have a dominant role. These findings highlight the crucial impact of previously logged user behavior data, which inherently captures patterns linked to individual Car IDs, reinforcing their importance in forecasting accuracy.

Fig. \ref{fig: importance2} illustrates the five base models used as features for the ensemble learning model, where their forecasts are frequently selected as influential features. The figure highlights that these features exhibit high selection frequency, with each contributing uniquely to the model's performance. Notably, LR has the highest importance in predicting connection duration, while XGBoost is most influential for energy demand. These results align with expectations, as the base models capture different aspects of the data, providing diverse perspectives that enhance the ensemble model’s predictive capability.

\begin{figure*}[!b]
\centerline{\includegraphics[width=0.5\linewidth]{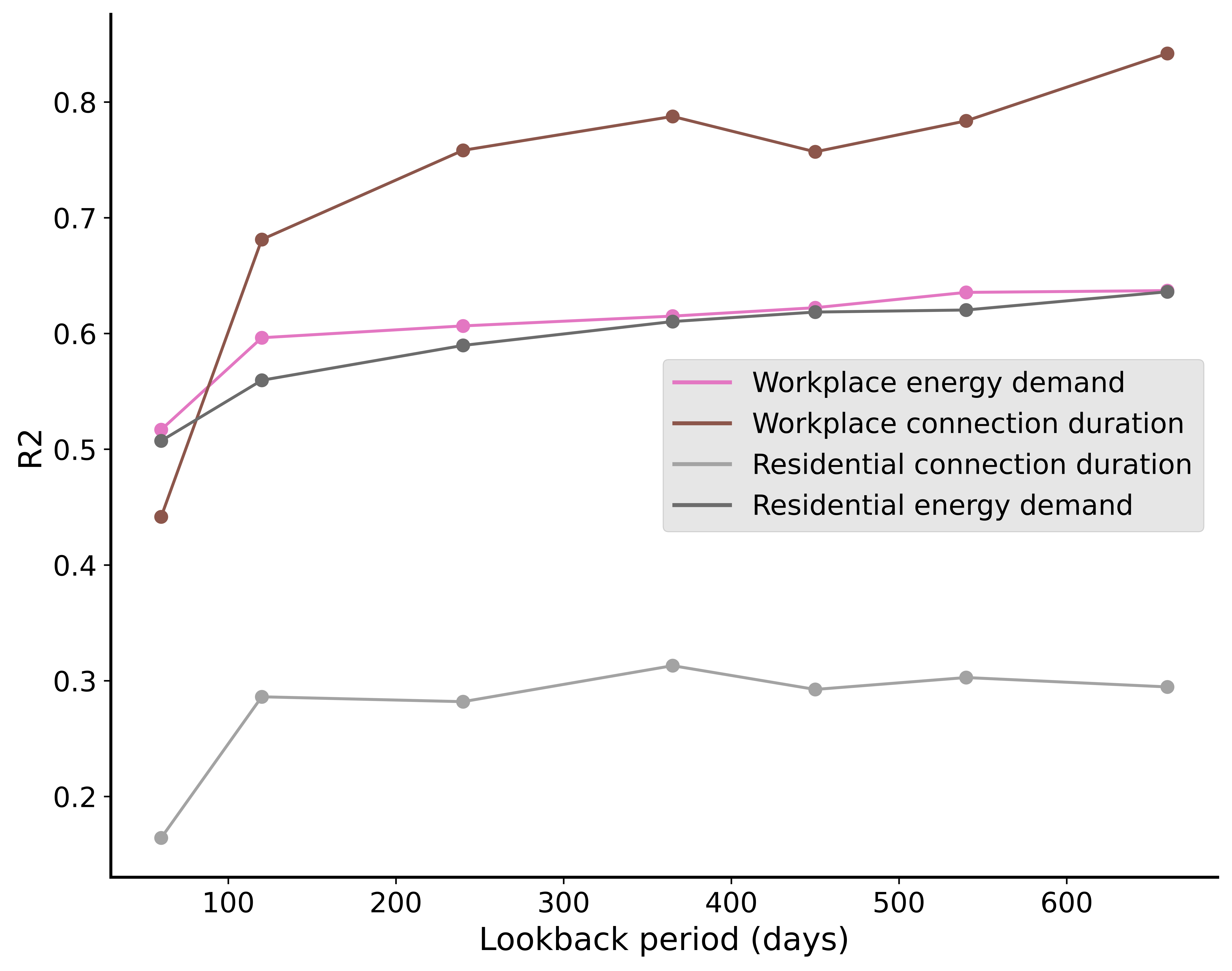}}
\caption{Impact of varying training dataset size on model accuracy for energy demand and connection duration forecasts across workplace and residential locations.}
\label{fig:lookback}
\end{figure*}

\subsection{Impact of lookback period and training size}
Fig. \ref{fig:lookback} illustrates the effect of training dataset size on model accuracy when implementing ensemble learning. The dataset size was expanded from 60 days to 660 days to evaluate its impact on accuracy for both workplace and residential EV charging parameters. A larger historical dataset enhances the model's ability to predict workplace energy demand patterns, as we observe a gradual increase in $R^2$ as the training window is expanded. However, for residential connections, we do not see much improvement with an increasing lookback period.

Furthermore, workplace connection duration accuracy improves with additional data, achieving the highest $R^2$ at the largest window size, suggesting that a more comprehensive dataset effectively captures complex temporal trends. Residential stations also benefit from a larger training window, with $R^2$ gradually increasing as more data is included. In contrast, residential connection duration exhibits more variability; although accuracy initially improves, the results fluctuate and are lower than in other cases.

\section{Discussion} \label{discussion}
This study proposes a forecasting approach for predicting individual EV charging parameters, focusing on a week-ahead forecast horizon over a one-year period. A comprehensive analysis of the proposed method’s effectiveness has been conducted using a real-world case study. To ensure that no data point or EV arrival was excluded, we implemented a dual-model approach, in contrast to previous studies that eliminated historical data or features associated with each car ID \cite{shahriar2021prediction} or eliminated first-time EV arrivals or restricted the dataset based on energy demand or connectivity duration thresholds \cite{shahriar2021prediction, kumar2024enhancing}. Our approach considers all EVs, with new arrivals using a model with limited historical features and previously seen EVs using a model with full historical data. The substantial performance gap between Model 1 and Model 2 underscores the necessity of incorporating historical data into EV charging forecasts. The poor predictive capability of Model 1 suggests that relying solely on generalized trends without car- and station-specific information is insufficient for accurate forecasting. Future work could explore alternative approaches, such as clustering similar charging behaviors, to enhance forecasts when historical records are unavailable.

The results demonstrate that even though more advanced models, particularly XGBoost and RF, provide a slight edge in forecasting energy demand and connection duration, simpler models like LR also perform remarkably well, especially in datasets with lower variability or straightforward patterns. This analysis also underscores that different models capture different aspects of charging behavior with different types of data points. Models like SVR and DT may excel on certain features or subsets, while XGBoost and RF perform better across the entire dataset by minimizing extreme errors. Given these complementary strengths, ensemble learning provides a strategic advantage by combining forecasts from multiple models, utilizing the strengths of each model on specific data points. This ensemble model optimally combines forecasts from all models to deliver more robust and accurate forecasts. With this approach, each model takes advantage of its specific strengths, ensuring higher accuracy even in cases where individual models may perform relatively poorly. In forecasting research, an accuracy improvement in this scale is significant, especially if external factors are complex and have high variability. The practical benefits of this ensemble model are especially relevant for CPOs, who require short-term or real-time forecasts to manage scheduling, pricing, and resource allocation. Grid operators also need accurate demand forecasts to minimize peak loads and grid strain.

It should be highlighted that a broad historical dataset is generally beneficial for model accuracy, but the degree of improvement varies by parameter and location. For example, workplace connection duration shows the greatest gains when a large historical dataset is used. However, residential connection duration patterns may not benefit as much from additional historical data due to their highly variable and potentially less predictable nature. Furthermore, feature importance analysis indicates that historical charging data, temporal features, and base model forecasts are consistently selected across training sessions as the most influential features. Accordingly, these core features capture the key patterns in EV charging parameters, while weather-related features contribute less.

In this study, historical data was used to forecast parameters for each EV plug-in event. However, the accurate modeling of EV user arrivals remains challenging due to the variability in user behavior and the influence of external factors, such as weather conditions, traffic patterns, or socio-economic variables. It is important that future research explores specialized methods, such as stochastic and probabilistic modeling, to better account for the variability of user arrival patterns.

Lastly, it is crucial to address the inherent uncertainty in forecasting EV charging parameters, particularly when these forecasts are used by CPOs in smart charging strategies or by grid operators for grid management. For reliable, real-world applications, uncertainty quantification is crucial since user behavior and external conditions often influence forecasts. For instance, conformal prediction \cite{renkema2024enhancing} can be used in future research to produce predictive intervals, which provide a range of likely outcomes rather than single-point forecasts.

\section{Conclusion} \label{conclusion}
This study developed an approach to predict two key EV charging parameters, i.e.,  energy demand and connection duration per session. Unlike previous studies, our approach leverages incremental learning and optimal feature and hyperparameter selection, integrating comprehensive feature data along with historical charging records to forecast week-ahead EV charging characteristics. An ensemble learning technique was used to enhance the accuracy of charging pattern forecasts using five widely used machine learning models. The proposed approach is highly adaptable and can be easily extended to different forecasting horizons or EV charging scenarios.

The ensemble learning strategy effectively combines the strengths of multiple machine learning models, demonstrating that while simpler models can provide relatively similar forecasts to more complex models, the ensemble model offers performance advantages by better capturing complex patterns and minimizing extreme errors. According to the feature selection analysis, historical and temporal features, as well as base model forecasts, played the greatest role in capturing EV charging patterns.  In addition, it can be concluded that larger training dataset sizes generally improve model accuracy. This underlines the value of extensive data for enhancing predictive performance. Overall, the proposed framework presents a versatile, data-driven, and highly scalable solution, contributing to the advancement of predictive modeling in EV charging infrastructure and energy management.

\section*{Acknowledgements}
This study is supported by the Dutch Ministry of Economic Affairs and Climate Policy and the Dutch Ministry of the Interior and Kingdom Relations through the ROBUST project under grant agreement MOOI32014 and by the European Union’s Horizon Europe Research and Innovation program through the 'SCALE - Smart Charging ALignment for Europe' project (grant agreement number 101056874). The authors would like to thank ROBUST consortium partners for fruitful discussions during the preparation of this paper.

\section*{Appendix. Hyperparameter sets} \label{app}
The range of considered hyperparameter sets for each considered forecasting model is presented in Table \ref{table:hyper}.

\begin{table}[b]
\centering
\caption{Tested hyperparameter set for the considered forecasting models for different EV charging parameters.} 
\label{table:hyper}
\begin{tabular}{p{5cm}|p{9cm}}
\hline
Forecasting model & Tested hyperparameter values \\ \hline
Multiple Linear Regression & No hyperparameters \\ 

Support Vector Regression & - Regularization parameter: [1, 10, 100, 1000] \newline
- Kernel coefficient: [0.01, 0.1, 1, 10] \newline
- Kernel type: ['rbf', 'linear']\newline
- Epsilon in the epsilon-tube: [0.01, 0.1, 0.5, 1]\\ 
                  
Decision Tree &  - Max. depth of tree: range(1, 20)\newline
- Min. samples to split: range(2, 20) \newline
- Min. samples per leaf: [1, 2, 4, 6] \newline
- Criterion: ['squared\_error', 'friedman\_mse', 'absolute\_error'] \\

Random Forest & - No. of trees in forest: [50, 75, 100, 150, 200, 250, 300]\newline
- Max. depth of tree: [2, 3, 4, 5, 6]\newline
- Min. no. of samples to split an internal node: [2, 3, 4, 5, 6]\newline
- Min. no. of samples to be at a leaf node: [1, 2, 3, 4, 5]\newline
- Considered no. of features: total no. of features \\
                  
Extreme Gradient Boosting & - No. of boosting round: [25, 50, 100, 150, 200]\newline
- Learning rate: [0.01, 0.1, 0.2]\newline
- Max. depth of tree: [2, 3, 4, 5]\newline
- Subsample ratio: [0.8, 0.9, 1.0]\newline
- Fraction of features per tree: [0.8, 0.9, 1.0]\newline
- Min. loss reduction for a split: [0, 0.1, 0.2, 0.5] \\

Ensemble Learning & Same hyperparameters as Extreme Gradient Boosting \\ \hline
\end{tabular}
\end{table}

\bibliographystyle{elsarticle-num} 
\bibliography{cas-refs}

\end{document}